\documentclass[aps,prd,a4paper,twocolumn]{revtex4}

\usepackage{graphicx}
\usepackage{bm}
\usepackage{amsfonts}
\usepackage{amsmath}

\usepackage[russian,ngerman,english]{babel}

\newcommand{\ve}[1]{\mbox{\boldmath$#1$}}

\let\oldbibitem\bibitem
\renewcommand\bibitem[2][]{\oldbibitem{#2}}

\begin{document}

\title{Time delay in the gravitational field of an axisymmetric body at rest with full mass and spin multipole structure}

\author{Sven Zschocke}

\affiliation{Institute of Planetary Geodesy - Lohrmann Observatory, TUD Dresden University of Technology,
Helmholtzstrasse 10, D-01069 Dresden, Germany}

\begin{abstract}
The time delay of a light signal which propagates in the gravitational field of an isolated body is considered. The body can be of arbitrary but time-independent shape and inner structure 
and can be in uniform rotational motion, while the center of mass of the body is assumed to be at rest. 
The gravitational field is given in the post-Newtonian scheme and in terms of the full set of mass-multipoles and spin-multipoles of the body. The asymptotic configuration is considered, where source and 
observer are located at spatial infinity from the massive body. It is found that in this asymptotic limit the higher multipole terms of time delay are related to the 
higher multipole terms of total light deflection. Furthermore, it is shown that the gauge terms vanish in this asymptotic configuration. 
In case of an axisymmetric body in uniform rotational motion, the higher multipole terms of time delay can be expressed in terms of Chebyshev polynomials. 
This fact allows one to determine the upper limits of the time delay for higher multipoles. These upper limits represent a criterion to identify those multipoles which  
contribute significantly to the time delay for a given accuracy of time measurements. It is found that the first mass-multipoles with $l \le 8$ and the first spin-multipoles 
with $l \le  3$ are sufficient for an accuracy on the femto-second scale of accuracy in time measurements. 
\end{abstract}

\pacs{95.10.Jk, 95.10.Ce, 95.30.Sf, 04.25.Nx, 04.80.Cc}

\begin{center}


\end{center}

\maketitle

\section{Introduction}\label{Section0}

In the classical Shapiro time delay \cite{Shapiro1} one considers the propagation of an electromagnetic signal, for instance a radar signal or a signal of visible light, 
in the gravitational field of a spherically symmetric body. Assume the space-time is covered by harmonic four-coordinates $x^{\mu} = \left(x^0, x^1, x^2, x^3\right)$, 
with time component $x^0$ and spatial components $x^1,x^2,x^3$ and the origin of the spatial coordinates is located at the center of mass of the body. 
Then, the light travel time of a signal, emitted at $\left(t_0, \ve{x}_0\right)$ and received at $\left(t_1, \ve{x}_1\right)$, is in the post-Newtonian (PN) scheme given by 
\cite{MTW,Book_Clifford_Will} 
\begin{eqnarray}
	\left(t_1 - t_0 \right) &=& \frac{R}{c} + \frac{2 G M}{c^3} \ln \frac{x_1 + \ve{\sigma} \cdot \ve{x}_1}{x_0 + \ve{\sigma} \cdot \ve{x}_0} + {\cal O}\left(c^{-3}\right)\;, 
        \label{Shapiro_1PN}
\end{eqnarray}

\noindent
where $M$ is the mass of the body and $\ve{\sigma}$ is the unit tangent vector along the light ray at minus infinity. The difference between the light travel time, 
$\left(t_1 - t_0 \right)$, and Euclidean distance, $R = \left|\ve{x}_1 - \ve{x}_0\right|$, divided by the speed of light, is the Shapiro time delay  
and belongs to the four classical tests of general relativity: perihelion precession of Mercury, light deflection at the Sun, gravitational redshift of light,
and light-travel time delay. The effect of time delay (\ref{Shapiro_1PN}) has been detected in $1968$ \cite{Shapiro2} and $1971$ \cite{Shapiro3}, 
which yields, for the round-trip Earth-Sun-Venus and Venus-Sun-Earth path, up to $251$ micro-seconds for radar signals grazing the Sun. The most accurate experiments of 
time delay measurements have been performed in $2003$ by using the Saturn orbiter {\it Cassini} as reflector, which amounts, for the round-trip Earth-Sun-Saturn and 
Saturn-Sun-Earth, up to $288$ micro-seconds, where a precision on the nano-second scale of accuracy has been achieved \cite{Shapiro6}.

Technological developments in time measurements by means of atomic clocks both on the ground as well as in space have made
giant progress during recent decades. Some remarks about todays atomic clock standards have been given
in the introductory section in \cite{Zschocke_Total_Light_Deflection_15PN}.
In particular, the accuracy of the standard deviation of up-to-date optical atomic clocks is $\Delta t / t = 10^{-19}$ \cite{Atomic_Clock1}. 
Such an accuracy corresponds to a precision of $0.001$ pico-seconds for a light signal which travels, for instance, from a giant planet of the Solar System towards
an observer located nearby the Earth. In fact, there are several mission proposals of the European Space Agency (ESA) \cite{Astrod1,Lator1,Odyssey,Sagas,TIPO,EGE}, 
aiming at time-delay measurements at the pico-second and sub-pico-second level of accuracy. 
Furthermore, the accuracy of time measurements improves by an order of magnitude every seven years \cite{Allan_Ashby}, and 
in near future the frequency standards will arrive a level of $\Delta t / t = 10^{-20} - 10^{-21}$, which corresponds to a precision of $0.01$ femto-seconds for a light signal 
from a giant planet of the Solar System towards an observer located nearby the Earth. 

In view of such rapid developments in time measurements, it becomes apparent that the mass-monopole approximation for Solar System bodies is not sufficient and 
a more realistic description of gravitational fields of Solar System bodies is necessary. The determination of the gravitational fields of  
Solar System bodies is achieved by decomposing the metric in terms of mass-multipoles $M_L$ (describe shape and inner structure of the massive body) 
and spin-multipoles $S_L$ (describe rotational motions and inner currents of the massive body) of these bodies. 
Then, the analytical formula for the time-delay becomes a complicated function of these multipoles \cite{Kopeikin1997},  
\begin{eqnarray}
	\left(t_1 - t_0 \right) &=& \frac{R}{c} + \sum\limits_{l=0}^{\infty} \Delta \tau_{\rm 1PN}^{M_L}\left(t_1,t_0\right) 
        + \sum\limits_{l=1}^{\infty} \Delta \tau_{\rm 1.5PN}^{S_L}\left(t_1,t_0\right) 
	\nonumber\\ 
	&& + {\cal O}\left(c^{-4} \right),
        \label{Shapiro_15PN_Introduction}
\end{eqnarray}

\noindent
where the first term in the sum $\left(l=0\right)$ is the mass-monopole term given by (\ref{Shapiro_1PN}).
In order to decide which multipoles are relevant for a given accuracy in time measurements, one has to determine the upper limits of the 
individual terms of the time delay formula in (\ref{Shapiro_15PN_Introduction}), that means the maximal absolute value of these terms.   

In a recent investigation it has been demonstrated that the effect of total light deflection is related to Chebyshev polynomials \cite{Zschocke_Total_Light_Deflection_15PN}. 
This fact has allowed for determining the upper limits of the total light deflection. As in case of total light deflection it is, therefore, the aim of this investigation to obtain 
an analytical formula for the multipole terms in (\ref{Shapiro_15PN_Introduction}) for the asymptotic configurations, where the source as well as the observer 
are located at spatial infinity from the massive body. It it found that in the asymptotic limit the multipole terms in (\ref{Shapiro_15PN_Introduction}) are related to 
Chebyshev polynomials, a fact that allows one to quantify the upper limits of absolute values of these terms to arbitrary multipole order. Using these criterions, one may easily 
decide, which terms in (\ref{Shapiro_15PN_Introduction}) are really relevant for a given goal accuracy in time delay measurements. 

The manuscript is organized as follows: In Section~\ref{Section1} the metric and the geodesic equation are considered. The Shapiro time delay is considered 
in Section~\ref{Section_Shapiro_1}. In Section~\ref{Relation} it is shown that in the asymptotic case, where source and observer are located infinitely  
far from the massive body, the total effect of time delay is related to the total light deflection. These results are applied in Section~\ref{Section_Shapiro_2} 
for the case of an axisymmetric body. A comparison with the literature and some numerical results of the time delay in the gravitational fields of Solar System bodies 
are given in Sections~\ref{Section_Comparison} and \ref{Section_Numerical_Values}. A summary is given in Section~\ref{Section_Summary}. 
The used notations are explained in Appendix~\ref{Appendix0}, while Appendices~\ref{Appendix1} and \ref{Appendix2} contain some further details of the calculations.

\section{Metric tensor and geodesic equation}\label{Section1}

\subsection{Metric tensor}

The curved space-time is described by a pair $\left({\cal M}, g_{\mu\nu}\right)$, where ${\cal M}$ is a four-dimensional differentiable manifold and $g_{\mu\nu}$ is the 
metric tensor of the manifold, and each point ${\cal P} \in {\cal M}$ represents a possible space-time event; we assume for the metric signature $\left(-,+,+,+\right)$.  
These ten components of the metric tensor $g_{\mu\nu}$ are determined by the ten field equations of gravity, 
which are valid in any coordinate system. The Bianchi identities reduce these field equations to only six independent equations. Therefore, 
four gauge conditions are imposed to fix the coordinates, which cover the physical manifold ${\cal M}$.  
We adopt harmonic four-coordinates, $z^{\mu} = \left(z^0, z^1, z^2, z^3\right)$, which are imposed by four harmonic gauge conditions \cite{Kopeikin_Efroimsky_Kaplan}  
\begin{eqnarray}
	\square_g \,z^{\mu} = 0\;, 
	\label{harmonic_gauge_1}
\end{eqnarray}

\noindent
where $\square_g = (-g)^{-1/2}\,\partial_{\mu}\,(-g)^{1/2}\,g^{\mu\nu}\,\partial_{\nu}$ is the general-covariant d'Alembert operator. 
If the gravitational fields are weak, then it is useful to decompose the metric tensor of the physical space-time \cite{MTW,Book_Clifford_Will,Kopeikin_Efroimsky_Kaplan,Carroll}, 
\begin{eqnarray}
	g_{\mu\nu}\left(t, \ve{x}\right) &=& \eta_{\mu\nu} + h_{\mu\nu}\left(t, \ve{x}\right), 
	\label{metric_A}
\end{eqnarray}

\noindent 
where $\eta_{\mu\nu} = \left(-1, +1, +1, +1\right)$ are the components of Minkowskian metric, while $h_{\mu\nu}$ are the metric perturbations which are small corrections to the 
Minkowskian metric: $\left|h_{\mu\nu}\right| \ll 1$. The decomposition (\ref{metric_A}) implies that the metric perturbations can be thought of as symmetric tensorial fields 
which propagate in the flat background space-time \cite{MTW,Kopeikin_Efroimsky_Kaplan,Carroll}. The flat space-time is described by a pair $\left({\cal M}_0, \eta_{\mu\nu}\right)$,
where ${\cal M}_0$ is the flat background manifold. The curved physical manifold ${\cal M}$ and the flat background manifold ${\cal M}_0$ are diffeomorphic to each other, which 
implicates a one-to-one correspondence between the points ${\cal P} \in {\cal M}$ and the points ${\cal Q} \in {\cal M}_0$. 

The flat background manifold ${\cal M}_0$ is assumed to be covered by harmonic four-coordinates $x^{\mu} = \left(x^0, x^1, x^2, x^3\right)$ which, 
according to (\ref{harmonic_gauge_1}), are imposed by the harmonic gauge condition,  
\begin{eqnarray}
        \square\,x^{\mu} = 0\;,
        \label{harmonic_gauge_2}
\end{eqnarray}

\noindent
where $\square = \eta^{\mu\nu}\,\partial_{\mu} \partial_{\nu}$ is the Lorentz-covariant d'Alembert operator. 
Then, by inserting the decomposition (\ref{metric_A}) into the exact field equations of gravity 
and keeping only terms linear in the metric perturbations, one obtains the linearized field equations of gravity, which in harmonic gauge read 
\begin{eqnarray}
	\square\,\overline{h}^{\mu\nu} &=& - \frac{16\,\pi\,G}{c^4}\, T^{\mu\nu}\;,
        \label{linearized_field_equations}
\end{eqnarray}

\noindent
where $T^{\mu\nu}$ is the stress-energy tensor of the source of matter. The following relations allow one to deduce the metric perturbations $h_{\mu\nu}$ 
from the metric density perturbations $\overline{h}_{\mu\nu}$ as soon as the solution of the linearized field equations (\ref{linearized_field_equations}) is found, 
\begin{eqnarray}
	\overline{h}_{\mu\nu} &=& h_{\mu\nu} - \frac{1}{2}\;h\;\eta_{\mu\nu}\;, 
	\label{perturbation_h_bar}
	\\ 
	h_{\mu\nu}  &=& \overline{h}_{\mu\nu} - \frac{1}{2}\;\overline{h}\;\eta_{\mu\nu}\;,
	\label{perturbation_h}
\end{eqnarray}

\noindent
with $h = h^{\alpha\beta}\,\eta_{\alpha\beta}$ and $\overline{h} = \overline{h}^{\alpha\beta}\,\eta_{\alpha\beta}$. 
These relations are valid at linear order in $h$. Furthermore, in linearized gravity the indices of tensors are lowered and raised by the Minkowskian metric. 

By imposing the Fock-Sommerfeld boundary condition \cite{Blanchet_Kopeikin_Schafer,KlionerKopeikin1992}, 
which implies the no-incoming radiation condition as well as the asymptotic flatness of space-time (isolated source of matter), the solution of the 
differential equation (\ref{linearized_field_equations}) is uniquely given by  
\begin{eqnarray}
	\overline{h}^{\mu\nu} = \frac{4\,\pi\,G}{c^4}\,\int_V d^3 x^{\prime}\,\frac{T^{\mu\nu}\left(t^{\prime},\ve{x}^{\prime}\right)}{\left|\ve{x} - \ve{x}^{\prime}\right|}\;,
        \label{retarded_solution}
\end{eqnarray}

\noindent
where the integral runs over the volume of the compact source of matter (body) and $t^{\prime} = t - c^{-1}\,\left|\ve{x} - \ve{x}^{\prime}\right|$. 

Without limiting generality, one may assume the harmonic four-coordinates $x^{\mu} = \left(x^0, x^1, x^2, x^3\right)$, which cover the flat background manifold ${\cal M}_0$, to be 
Cartesian. However, from the harmonic gauge condition (\ref{harmonic_gauge_2}) one concludes, that the harmonic gauge does not fix these harmonic four-coordinates $x^{\mu}$ uniquely, 
but allows for smooth deformations \cite{MTW,Kopeikin_Efroimsky_Kaplan}  
\begin{equation}
        x^{\alpha}_{{\rm can}} = x^{\alpha} + w^{\alpha}(x^{\beta}) ,
\label{coordinate_transformation}
\end{equation}

\noindent
if these gauge vector fields $w^{\alpha}$ satisfy $\square\,w^{\alpha} = 0$. The label of these new coordinates $\{x_{\rm can}\}$ abbreviates the term {\it canonical}. 
It is emphasized that the gauge transformations in (\ref{coordinate_transformation}) are tending to $0$ at spatial infinity and they are small, $|w^{\alpha}| \ll |x^{\alpha}|$, 
and in the sense that the derivatives of the gauge functions $w^{\alpha}$ with respect to space and time are of the same order as the metric perturbations, 
$w^{\alpha}_{\,,\,\mu} = {\cal O}\left(h^{\alpha}_{\mu}\right)$, hence $|w^{\alpha}_{\,,\,\mu} | \ll 1$. 
The residual gauge transformation (\ref{coordinate_transformation}) implies a residual gauge transformation of the metric tensor, 
\begin{equation}
        g_{\alpha\beta}\left(t,\ve{x}\right)
        = \frac{\partial x^{\mu}_{\rm can}}{\partial x^{\alpha}}\, \frac{\partial x^{\nu}_{\rm can}}{\partial x^{\beta}}\,
        g_{\mu\nu}^{\rm can}\left(t_{\rm can} , \ve{x}_{\rm can}\right). 
\label{transformation_metric_tensor_A1}
\end{equation}

\noindent
By inserting (\ref{coordinate_transformation}) into (\ref{transformation_metric_tensor_A1}) and performing a series expansion
of the metric tensor on the right-hand side around the old (Cartesian) coordinates $\{x\}$, one obtains
($\partial_{\alpha}\,f \equiv f_{\,,\,\alpha} \equiv \partial f/ \partial x^{\alpha}$):
\begin{equation}
        g_{\alpha\beta} \left(t,\ve{x}\right) = g_{\alpha\beta}^{\rm can} \left(t,\ve{x}\right) + \partial_{\alpha} w_{\beta} \left(t,\ve{x}\right)
        + \partial_{\beta} w_{\alpha}\left(t,\ve{x}\right).
\label{transformation_metric_tensor_A2}
\end{equation}

\noindent
The unique solution in (\ref{retarded_solution}) can be expressed in terms of six Cartesian symmetric and trace-free (STF) multipoles $\hat{F}_L$ (Eq.~(8.4) in \cite{Thorne} 
or Eqs.~(5.3a) - (5.3c) in \cite{Multipole_Damour_2}), 
\begin{equation}
	\overline{h}^{\mu\nu}\left(t, \ve{x}\right) = \frac{4 \,G}{c^4} \sum\limits_{l=0}^{\infty} \frac{\left(-1\right)^l}{l!}\,\hat{\partial}_L 
	\left[\frac{\hat{F}_L^{\mu\nu}\left(u\right)}{r}\right]\,,
	\label{unique_solution_STF}
\end{equation}

\noindent
where $u = t - c^{-1}\,r$ is the retarded time, $r = \left|\ve{x}\right|$, the STF multipoles $\hat{F}_L$ are given by Eqs.~(5.4a) - (5.4c) in \cite{Multipole_Damour_2}, and
\begin{equation}
        \hat{\partial}_L = {\rm STF}_{i_1 \dots i_l}\,\frac{\partial}{\partial x^{i_1}} \dots \frac{\partial}{\partial x^{i_l}}\,,
\label{Differential_Operator}
\end{equation}

\noindent
where the {\it hat} in $\hat{F}_L$ and in $\hat{\partial}_L$ indicates STF operation with respect to the spatial components of the multi-index $L = i_1 \dots i_l$. 
A detailed proof of this theorem in (\ref{unique_solution_STF}) has been presented in \cite{Zschocke_Theorem}. 

The solution in (\ref{unique_solution_STF}) is unique and represents the most general solution of the linearized gravity for an isolated compact source of matter. By extensive use of 
STF Cartesian tensor techniques, it has been demonstrated in \cite{Thorne,Blanchet_Damour1,Multipole_Damour_2} that the general solution in (\ref{unique_solution_STF}) can be 
written in terms of six STF multipoles $\{\hat{M}_L, \hat{S}_L, \hat{W}_L, \hat{X}_L, \hat{Y}_L, \hat{Z}_L\}$ \cite{Thorne,Blanchet_Damour1,Multipole_Damour_2}. Accordingly, the 
metric tensor can be written in the following form, 
\begin{eqnarray}
	h_{\alpha\beta}\left(t,\ve{x}\right) &=& h^{{\rm can}}_{\alpha\beta} \left[\hat{M}_L, \hat{S}_L\right]
	+ \partial_{\alpha} w_{\beta}\left[\hat{W}_L, \hat{X}_L, \hat{Y}_L, \hat{Z}_L\right]
	\nonumber\\ 
	&& + \partial_{\beta} w_{\alpha}\left[\hat{W}_L, \hat{X}_L, \hat{Y}_L, \hat{Z}_L\right].
        \label{PN_Expansion_1PN_15PN}
\end{eqnarray}

\noindent 
The canonical part of the metric perturbations in (\ref{PN_Expansion_1PN_15PN}) depends on two multipoles only, namely mass-multipoles and spin-multipoles $\{\hat{M}_L, \hat{S}_L\}$,  
while the gauge terms of the metric perturbation in (\ref{PN_Expansion_1PN_15PN}) depends on four multipoles, $\{\hat{W}_L, \hat{X}_L, \hat{Y}_L, \hat{Z}_L\}$. 
The form of the metric perturbations in (\ref{PN_Expansion_1PN_15PN}) depends on the chosen coordinate system. For instance, 
in canonical coordinates there are no gauge terms at all and only the canonical term $h^{{\rm can}}_{\alpha\beta}$ would remain. Here, we will use this most general form 
in (\ref{PN_Expansion_1PN_15PN}) in order to demonstrate that these gauge terms vanish at spatial infinity. 

We will consider the metric of bodies with time-independent multipoles and where the center of mass of the body is assumed to be at rest with respect to the harmonic coordinate system. 
Then, the canonical metric perturbations in (\ref{PN_Expansion_1PN_15PN}) are separated into two pieces, 
$h^{\rm can}_{\alpha\beta} = h_{\alpha\beta}^{\left(2\right)\,{\rm can}} + h_{\alpha\beta}^{\left(3\right)\,{\rm can}}$, 
\begin{eqnarray}
        h_{00}^{\left(2\right)\,{\rm can}} &=& - \frac{2 G}{c^2} \sum\limits_{l=0}^{\infty} \frac{\left(-1\right)^l}{l!}\,
        \hat{\partial}_L \frac{\hat{M}_L}{r} \;,
        \label{Metric1}
        \\ 
        h_{0i}^{\left(3\right)\,{\rm can}} &=& + \frac{4 G}{c^3} \sum\limits_{l=1}^{\infty} \frac{\left(-1\right)^l\,l}{\left(l+1\right)!} \,
        \epsilon_{iab}\,\hat{\partial}_{a L-1} \frac{\hat{S}_{b L-1}}{r}\;,
        \label{Metric2}
\end{eqnarray}

\noindent
while $h_{ij}^{\left(2\right)\,{\rm can}}  = h_{00}^{\left(2\right)\,{\rm can}}\,\delta_{ij}$. 
The derivatives $\hat{\partial}_L$ are not acting on the multipoles, because they are independent of space and time here. 
These mass-multipoles and spin-multipoles in (\ref{Metric1}) and  (\ref{Metric2}) in the stationary case, that means in case of a time-independent source of matter, are given by 
\begin{eqnarray}
	\hat{M}_L &=& \int d^3 x\,\hat{x}_L\,\Sigma\;, 
        \label{Mass_Multipoles}
	\\
	\hat{S}_L &=& \int d^3 x\, \epsilon_{jk<i_l}\,\hat{x}_{L-1>} \,x^j\,\Sigma^k\;,
	\label{Spin_Multipoles}
\end{eqnarray}

\noindent 
where $\Sigma = \left(T^{00} + T^{kk}\right)/c^2$ and $\Sigma^k = T^{0k}/c$ with the stress-energy tensor $T^{\mu\nu}$ of the source of matter (cf. Eq.~(\ref{linearized_field_equations})), 
and the integration runs over the volume of the body. The time-independence of the multipoles implies time-independence of the metric. 

The gauge functions in (\ref{PN_Expansion_1PN_15PN}) have been determined in \cite{Thorne,Blanchet_Damour1,Multipole_Damour_2,Multipole_Blanchet_1} and read:
\begin{eqnarray}
	w^{0} &=& + \frac{4 G}{c^3} \sum\limits_{l=0}^{\infty} \frac{\left(-1\right)^l}{l!}\,\hat{\partial}_L \frac{\hat{W}_L}{r} \;,
        \label{gaugefunction1}
	\\
	w^{i} &=& - \frac{4 G}{c^2} \sum\limits_{l=0}^{\infty} \frac{\left(-1\right)^l}{l!}\,\hat{\partial}_{i L} \frac{\hat{X}_L}{r} 
	- \frac{4 G}{c^2} \sum\limits_{l=1}^{\infty} \frac{\left(-1\right)^l}{l!}\,\hat{\partial}_{L - 1} \frac{\hat{Y}_{i L - 1}}{r}
	\nonumber\\ 
	&& - \frac{4 G}{c^2} \sum\limits_{l=1}^{\infty} \frac{\left(-1\right)^l}{l!}\,\frac{l}{l + 1}\, 
	\epsilon_{i ab} \,\hat{\partial}_{a L - 1} \frac{\hat{Z}_{b L - 1}}{r} \,.
        \label{gaugefunction2}
\end{eqnarray}

\noindent
As mentioned above, these gauge vectors in (\ref{gaugefunction1}) - (\ref{gaugefunction2}) represent the most general form of the gauge terms in the metric tensor. The multipoles 
$\hat{W}_L,\hat{X}_L,\hat{Y}_L,\hat{Z}_L$ of the gauge functions (\ref{gaugefunction1}) and (\ref{gaugefunction2}) have been determined in \cite{Multipole_Damour_2,Multipole_Blanchet_1} 
and read in the stationary case as follows,  
\begin{eqnarray}
	\hat{W}_L &=& + \frac{2 l + 1}{\left(l + 1\right) \left(2 l + 3\right)} \int d^3 x\,\hat{x}_{Lj}\,\Sigma^j\;, 
        \label{Multipoles_W_L}
        \\
	\hat{X}_L &=& + \frac{2 l + 1}{\left(2 l + 2\right) \left(l + 2\right) \left(2 l + 5\right)} \int d^3 x\,\hat{x}_{j k L}\,\Sigma^{jk}\;,
        \label{Multipoles_X_L}
        \\
	\hat{Y}_L &=& - \int d^3 x\,\hat{x}_L\,\Sigma^{kk}\;,
        \label{Multipoles_Y_L}
        \\
	\hat{Z}_L &=& - \frac{2 l + 1}{\left(l + 2\right) \left(2 l + 3\right)} \!\int d^3 x \,\epsilon_{j k <i_l} \hat{x}_{L - 1 > k m} \Sigma^{j m},
        \label{Multipoles_Z_L}
\end{eqnarray}

\noindent
where $\Sigma^{jk} = T^{jk}/c^2$, with $T^{jk}$ being the spatial components of the stress-energy tensor of the body, 
and the integration runs over the volume of the body.  

We will show that the gauge terms in (\ref{PN_Expansion_1PN_15PN}) have no impact on the Shapiro time delay in case of infinite distance of source and observer from the massive body 
(see Appendix~\ref{Appendix1}). This specific case reflects the general fact that $g_{\alpha\beta}$ and $g_{\alpha\beta}^{\rm can}$ in (\ref{transformation_metric_tensor_A2})
are physically equivalent, that means they lead to same observables. This statement is valid in case the gauge transformations (\ref{coordinate_transformation}) are small 
and are tending to $0$ at spatial infinity, as emphasized in the text below Eq.~(\ref{coordinate_transformation}).

\subsection{Geodesic equation}

In flat Minkowskian space-time, assumed to be covered by Cartesian four-coordinates, 
a light signal, emitted by a source at $\left(t_0, \ve{x}_0\right)$ in some direction specified by a unit vector $\ve{\sigma}$, 
propagates along a straight line, given by 
\begin{eqnarray}
	\ve{x}_{\rm N} &=& \ve{x}_0 + c \left( t - t_0\right) \ve{\sigma}\;,  
	\label{Unperturbed_Lightray_1}
\end{eqnarray}

\noindent
and its absolute value is 
\begin{eqnarray}
        r_{\rm N} = \sqrt{\left(x_0\right)^2 + 2\,\ve{\sigma} \cdot \ve{x}_0 \;c \left(t - t_0\right) + c^2 \left(t-t_0\right)^2}\,,
\label{Unperturbed_Lightray_Absolute_1}
\end{eqnarray}

\noindent 
where the index N in (\ref{Unperturbed_Lightray_1}) and (\ref{Unperturbed_Lightray_Absolute_1}) stands for Newtonian. 
In what follows we also need the so-called impact vector of the unperturbed light ray,
\begin{eqnarray}
	\ve{d}_{\sigma} &=& \ve{\sigma} \times \left(\ve{x}_{\rm N} \times \ve{\sigma}\right), 
        \label{impact_vector_d_sigma}
\end{eqnarray}

\noindent
with its absolute value, the impact parameter of the unperturbed light ray, 
\begin{eqnarray}
	d_{\sigma} = \left|\ve{\sigma} \times \ve{x}_{\rm N}\right|. 
        \label{absolute_value_impact_vector_d_sigma}
\end{eqnarray}

\begin{figure}[!ht]
\includegraphics[scale=0.12]{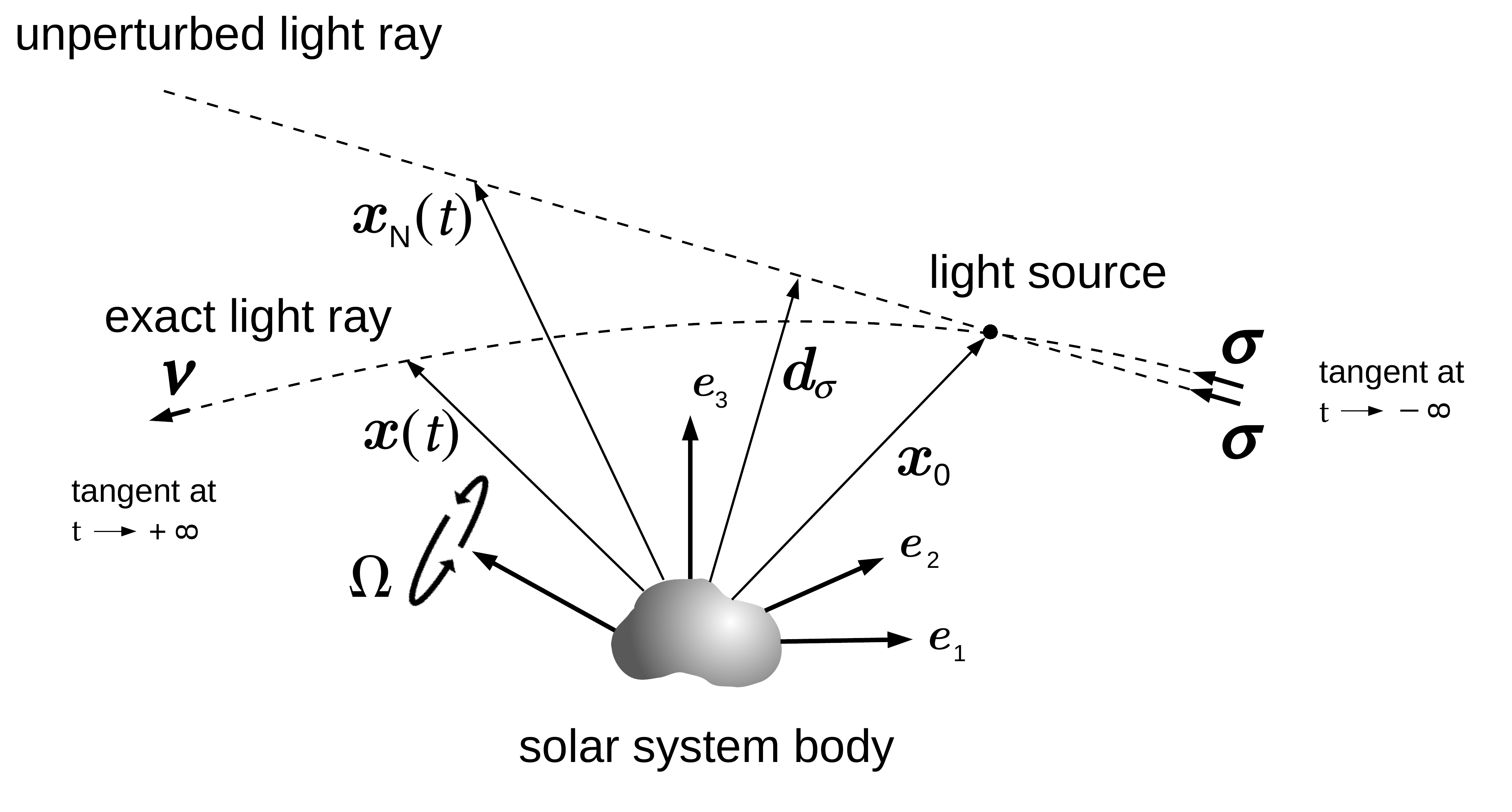}
        \caption{A geometrical representation of the propagation of a light signal through the gravitational field of a massive Solar System body
        at rest. The axes of inertia are denoted by $\ve{e}_1, \ve{e}_2, \ve{e}_3$. 
	In stationary case, the body is of arbitrary but time-independent shape and inner structure and 
	the vector of angular velocity $\ve{\Omega}$ and its absolute value $\Omega$ are time-independent. 
	The light signal is emitted by the light source at $\ve{x}_0$ and propagates along the exact light trajectory $\ve{x}\left(t\right)$.
	The unit tangent vectors along the light trajectory at minus and plus infinity are $\ve{\sigma}$ and $\ve{\nu}$.
        The unperturbed light ray $\ve{x}_{\rm N}\left(t\right)$ is given by Eq.~(\ref{Unperturbed_Lightray_1}) and propagates in the direction
        of $\ve{\sigma}$ along a straight line through the position of the light source at $\ve{x}_0$. The impact vector $\ve{d}_{\sigma}$ of the
        unperturbed light ray is given by Eq.~(\ref{impact_vector_d_sigma}).}
\label{Diagram1}
\end{figure}

\noindent 
The impact vector (\ref{impact_vector_d_sigma}) points from the center of mass of the body towards the unperturbed light ray at its closest encounter; 
for a graphical elucidation see Fig.~\ref{Diagram1}. 

In curved space-time, $\left({\cal M}, g_{\mu\nu}\right)$, the trajectory of a light signal is determined by the geodesic equation, 
which in terms of coordinate time reads as follows \cite{Brumberg1991,Kopeikin_Efroimsky_Kaplan,MTW}
\begin{eqnarray}
        && \frac{\ddot{x}^{i}\left(t\right)}{c^2} + \Gamma^{i}_{\mu\nu} \frac{\dot{x}^{\mu}\left(t\right)}{c} \frac{\dot{x}^{\nu}\left(t\right)}{c}
- \Gamma^{0}_{\mu\nu} \frac{\dot{x}^{\mu}\left(t\right)}{c} \frac{\dot{x}^{\nu}\left(t\right)}{c} \frac{\dot{x}^{i}\left(t\right)}{c} = 0\;,
\nonumber\\ 
        \label{Geodetic_Equation2}
\end{eqnarray}

\noindent
where a dot denotes total derivative with respect to coordinate time, and $\Gamma^{\alpha}_{\mu\nu}$ are the Christoffel symbols, given by  
\cite{Kopeikin_Efroimsky_Kaplan,Brumberg1991,MTW}   
\begin{eqnarray}
\Gamma^{\alpha}_{\mu\nu} = \frac{1}{2}\,g^{\alpha\beta} 
\left(\frac{\partial g_{\beta\mu}}{\partial x^{\nu}} + \frac{\partial g_{\beta\nu}}{\partial x^{\mu}} 
- \frac{\partial g_{\mu\nu}}{\partial x^{\beta}}\right).   
\label{Christoffel_Symbols2}
\end{eqnarray}

\noindent
The geodesic equation is a differential equation of second order of one variable, $t$, thus a unique solution of (\ref{Geodetic_Equation2}) 
necessitates two initial-boundary conditions: the spatial position of light source $\ve{x}_0$ and the unit-direction $\ve{\sigma}$ of the 
light signal at minus infinity \cite{Brumberg1991,Kopeikin1997,KopeikinSchaefer1999_Gwinn_Eubanks,KlionerKopeikin1992,Zschocke_1PN,Zschocke_15PN}: 
\begin{eqnarray}
        \ve{\sigma} &=& \frac{\dot{\ve{x}}\left(t\right)}{c}\,\bigg|_{t = - \infty} \quad {\rm with} \quad 
        \ve{\sigma} \cdot \ve{\sigma} = 1\;,  
        \label{Initial_A}
        \\ 
        \ve{x}_0 \; &=& \; \ve{x}\left(t\right) \bigg|_{t = t_0}\;.  
        \label{Initial_B}
\end{eqnarray}

\noindent
By using the initial-boundary conditions (\ref{Initial_A}) and by inserting the decomposition (\ref{metric_A}) into (\ref{Geodetic_Equation2}), 
which implies weak gravitational fields, the solution of the second integration of geodesic equation (\ref{Geodetic_Equation2}) is given by 
\begin{eqnarray}
        \ve{x}\left(t\right) &=& \ve{x}_0 + c \left(t-t_0\right) \ve{\sigma} + \Delta \ve{x}\left(t,t_0\right), 
        \label{Introduction_2}
\end{eqnarray}

\noindent
where $\Delta \ve{x}$ are small corrections to the trajectory of the unperturbed light ray (\ref{Unperturbed_Lightray_1}). Eq.~(\ref{Introduction_2}) represents the trajectory of a 
light signal propagating in the flat background space-time $\left({\cal M}_0, \eta_{\mu\nu}\right)$; for a graphical elucidation see Fig.~\ref{Diagram1}.
The solution of the initial value problem (\ref{Introduction_2}) implies the following limit,
\begin{eqnarray}
        \lim_{t \rightarrow t_0} \Delta \ve{x}\left(t,t_0\right) &=& 0 \;, 
        \label{Introduction_4} 
\end{eqnarray}

\noindent
in order to be consistent with the condition  (\ref{Initial_B}).

The geodesic equation in 1.5PN approximation can be deduced from the exact geodesic equation (\ref{Geodetic_Equation2}) and is given by \cite{Brumberg1991}  
\begin{eqnarray}
	\frac{\ddot{x}^i \left(t\right)}{c^2} &=& \frac{1}{2} \,h_{00,i} - h_{00,j}\,\sigma^i \sigma^j - h_{ij,k}\, \sigma^j \sigma^k + \frac{1}{2}\,h_{jk,i}\,\sigma^j\sigma^k
\nonumber\\ 
	&& \hspace{-0.75cm} - h_{0i,j}\,\sigma^j + h_{0j,i}\,\sigma^j - h_{0j,k}\,\sigma^i\,\sigma^j \sigma^k + {\cal O}\left(c^{-4}\right)\;, 
\label{geodesic_equation_5}
\end{eqnarray}

\noindent
where we have omitted all those terms which contain a derivative of the metric perturbations with respect to time, because we consider the stationary case, 
that is the case of time-independent metric. Note, that in stationary case the geodesic equation in 1.5PN approximation of the post-Newtonian (PN) scheme and 
the geodesic equation in 1PM approximation of the post-Minkowskian (PM) scheme agree with each other \cite{Klioner_Peip}. 
By inserting the metric perturbation (\ref{PN_Expansion_1PN_15PN}) into the geodesic equation (\ref{geodesic_equation_5}),
one may separate the geodesic equation into a canonical term, $\ddot{\ve{x}}_{\rm can}$, plus a gauge term, $\ddot{\ve{x}}_{\rm gauge}$, as follows:
\begin{equation}
	\frac{\ddot{\ve{x}} \left(t\right)}{c^2} = \frac{\ddot{\ve{x}}_{\rm can} \left(t\right)}{c^2} + \frac{\ddot{\ve{x}}_{\rm gauge} \left(t\right)}{c^2} + {\cal O}\left(c^{-4}\right), 
\label{geodesic_equation_10}
\end{equation}

\noindent
where the spatial components of these terms are
\begin{eqnarray}
\frac{\ddot{x}_{\rm can}^i \left(t\right)}{c^2} &=& h^{\left(2\right)\,{\rm can}}_{00,i} - 2\,h^{\left(2\right)\,{\rm can}}_{00,j}\,\sigma^i \sigma^j 
- h^{\left(3\right)\,{\rm can}}_{0i,j}\,\sigma^j 
\nonumber\\ 
	&& + h^{\left(3\right)\,{\rm can}}_{0j,i}\,\sigma^j - h^{\left(3\right)\,{\rm can}}_{0j,k}\,\sigma^i\,\sigma^j \sigma^k\;, 
\label{geodesic_equation_can}
\\
        \frac{\ddot{x}_{\rm gauge}^i \left(t\right)}{c^2} &=& \partial_j\,w^0_{\,,\,k} \,\sigma^i \sigma^j \sigma^k - \partial_j\,w^i_{\,,\,k} \,\sigma^j \sigma^k \;. 
\label{geodesic_equation_gauge}
\end{eqnarray}

\noindent
The metric perturbations in (\ref{geodesic_equation_can}) are given by (\ref{Metric1}) and (\ref{Metric2}), while 
the gauge functions in (\ref{geodesic_equation_gauge}) are given by (\ref{gaugefunction1}) and (\ref{gaugefunction2}). 
The metric tensor with the metric perturbations (\ref{PN_Expansion_1PN_15PN}) is valid in the entire space-time, while in the geodesic equation (\ref{geodesic_equation_10})  
with (\ref{geodesic_equation_can}) and (\ref{geodesic_equation_gauge}), the arguments of the metric perturbations are taken along the light trajectory, that means they have 
to be replaced by $\ve{x} = \ve{x}_{\rm N} + {\cal O}(c^{-2})$ and $r = |\ve{x}_{\rm N}| + {\cal O}(c^{-2})$. 

The integration of (\ref{geodesic_equation_10}) yields the light trajectory in 1.5PN approximation, formally given by  
\begin{equation}
\ve{x}\left(t\right) = \ve{x}_0 + c \left(t - t_0\right) \ve{\sigma} + \Delta \ve{x}_{\rm can}\left(t,t_0\right) + \Delta \ve{x}_{\rm gauge}\left(t,t_0\right). 
\label{geodesic_equation_15}
\end{equation}

\noindent 
In particular, one has to insert the multipole decomposition of the metric perturbations (\ref{PN_Expansion_1PN_15PN}) into the geodesic equation (\ref{geodesic_equation_5}) 
and then one has to apply advanced integration methods developed in \cite{Kopeikin1997}, which yields 
\begin{eqnarray}
	\ve{x}\left(t\right) &=& \ve{x}_0 + c \left(t - t_0\right) \ve{\sigma} 
        \nonumber\\ 
	&& + \sum\limits_{l=0}^{\infty} \Delta \ve{x}_{\rm 1PN}^{M_L}\left(t, t_0\right) + \sum\limits_{l=1}^{\infty} \Delta \ve{x}_{\rm 1.5PN}^{S_L}\left(t, t_0\right) 
	\nonumber\\
	&& +\, \Delta \ve{x}_{\rm gauge} \left(t, t_0\right).  
	\label{Light_Trajectory_PN} 
\end{eqnarray}

\noindent 
The first line describes the straight trajectory of the unperturbed light ray, the second line represents the perturbations to the unperturbed light ray caused by the canonical terms 
of the metric tensor, and the third line are the perturbations to the unperturbed light ray caused by the gauge terms of the metric tensor. The canonical terms 
$\Delta \ve{x}_{\rm 1PN}^{M_L}$ and $\Delta \ve{x}_{\rm 1.5PN}^{S_L}$ have been obtained at the very first time by advanced integration methods in \cite{Kopeikin1997}, 
while the gauge term $\Delta \ve{x}_{\rm gauge}$ has been calculated in Appendix~\ref{Appendix1} by the same approach developed in \cite{Kopeikin1997}.

\section{Time delay in field of an arbitrary body}\label{Section_Shapiro_1}

In \cite{Kopeikin1997} advanced integration methods have been introduced that one allow to integrate (\ref{geodesic_equation_5}) exactly.  
The basic ideas of the conception has originally been introduced in \cite{Kopeikin1997} for bodies with time-independent multipoles and where 
the center of mass of the body is assumed to be at rest with respect to the harmonic coordinate system. This approach has further been developed for the case 
of light propagation in the gravitational field of a time-dependent source of matter, where the center of mass of the body was also assumed to be at rest 
with respect to the harmonic coordinate system \cite{KopeikinSchaefer1999_Gwinn_Eubanks,KopeikinKorobkovPolnarev2006,KopeikinKorobkov2005}. Later, these mathematical tools 
developed in \cite{Kopeikin1997,KopeikinSchaefer1999_Gwinn_Eubanks,KopeikinKorobkovPolnarev2006,KopeikinKorobkov2005} have been applied to the case of light propagation 
in the gravitational fields of $N$ slowly moving bodies with time-dependent multipoles \cite{Zschocke_1PN,Zschocke_15PN}.

In the approach in \cite{Kopeikin1997} two new parameters were introduced,
\begin{eqnarray}
	c \tau &=& \ve{\sigma} \cdot \ve{x}_{\rm N}\;,
	\label{Parameter1}
	\\
	\xi^i &=& P^i_j\,x_{\rm N}^j\;,  
	\label{Parameter2}
\end{eqnarray}
 
\noindent
where $P^{ij} = \delta^{ij} - \sigma^i \sigma^j $ is a projection operator onto the plane perpendicular to vector $\ve{\sigma}$. 
The unperturbed light ray (\ref{Unperturbed_Lightray_1}) expressed in terms of these new variables takes the form
\begin{eqnarray}
        \ve{x}_{\rm N} &=& \ve{\xi} + c \tau\,\ve{\sigma}\;,
        \label{Unperturbed_Lightray_2}
\end{eqnarray}

\noindent 
with its absolute value 
\begin{eqnarray}
	r_{\rm N} = \sqrt{\xi^2 + c^2 \tau^2}\;.
        \label{Unperturbed_Lightray_Absolute_2}
\end{eqnarray}

\noindent
In favor of a simpler notation, we will not introduce new notations for the unperturbed light ray (\ref{Unperturbed_Lightray_1}) given in terms of the standard variables 
$\left(c t, \ve{x}\right)$ and for the unperturbed light ray (\ref{Unperturbed_Lightray_2}) given in terms of the auxiliary variables $\left(c \tau, \ve{\xi}\right)$. Similarly, 
the same notation will be used for their absolute values in (\ref{Unperturbed_Lightray_Absolute_1}) and (\ref{Unperturbed_Lightray_Absolute_2}). 

The three-vector $\ve{\xi}$ in (\ref{Parameter2}) actually coincides with the impact vector defined by (\ref{impact_vector_d_sigma}). The use of two different 
notations for the same vector is appropriate for the following reason: the three-vector $\ve{\xi}$ is laying in the two-dimensional plane perpendicular to $\ve{\sigma}$, 
hence only two components are independent, which implies $\partial \xi^i / \partial \xi^j = P^i_j$. However, in practical calculations it is convenient to treat the 
spatial components of this vector as formally independent, which implies $\partial \xi^i / \partial \xi^j = \delta^i_j$. Therefore, a subsequent projection onto this 
two-dimensional plane by means of $P^{ij}$ is necessary \cite{KopeikinSchaefer1999_Gwinn_Eubanks,Book_Soffel_Han}.  
That is why two different notations for \ve{\xi} and $\ve{d}_{\sigma}$ are in use here. Then, for a spatial derivative expressed in terms of these new variables, one obtains  
\begin{eqnarray}
	\frac{\partial}{\partial x^i} &=& P_i^j\,\frac{\partial}{\partial \xi^j} + \sigma_i\,\frac{\partial}{\partial c \tau}\;. 
	\label{spatial_derivative}
\end{eqnarray}

\noindent
Using (\ref{spatial_derivative}) and the binomial theorem, one finds the differential operator in (\ref{Differential_Operator}) 
expressed in terms of these new variables,  
\begin{eqnarray}
	\widehat{\partial}_{L} &=& {\rm STF}_{i_1 \dots i_l}\;\sum\limits_{p=0}^{l} \frac{l!}{\left(l-p\right)!\;p!}\;\sigma_{i_1}\,...\,\sigma_{i_p}  
	\, P_{i_{p+1}}^{j_{p+1}}\;...\;P_{i_l}^{j_l}
	\nonumber\\ 
	&& \times \frac{\partial}{\partial \xi^{j_{p+1}}}\;...\;
\frac{\partial}{\partial \xi^{j_{l}}}\;\left(\frac{\partial}{\partial c\tau}\right)^p \,,  
\label{Transformation_Derivative_3}
\end{eqnarray}

\noindent
where the symbol {\it widehat} indicates STF operation with respect to the spatial indices $i_1 \dots i_l$. 
Here we prefer to use the operator as given by Eq.~(\ref{Transformation_Derivative_3}) where $\partial \xi^i / \partial \xi^j = \delta^i_j$, 
while if one applies the operator as given by Eq.~(24) in \cite{Kopeikin1997} then 
$\partial \xi^i / \partial \xi^j = P^i_j$. The final results of either these operations are identical. 
Then, using the basic integral (25) in \cite{Kopeikin1997}, one finds for the second integration the formula given by Eq.~(27) 
in \cite{Kopeikin1997}, which leads to the solution for the second integration of geodesic equation (\ref{geodesic_equation_5}). 

From the solution for the light trajectory in (\ref{Light_Trajectory_PN}), one obtains the time-of-flight in the gravitational field of a body at rest 
with full mass-multipole and spin-multipole structure, given by the following formula \cite{Kopeikin1997},  
\begin{eqnarray}
	\left(t_1 - t_0 \right) &=& \frac{R}{c} + \sum\limits_{l=0}^{\infty} \Delta \tau_{\rm 1PN}^{M_L}\!\left(t_1,t_0\right)  
	+ \sum\limits_{l=1}^{\infty} \Delta \tau_{\rm 1.5PN}^{S_L}\!\left(t_1,t_0\right) 
	\nonumber\\ 
	&& +\,\Delta \tau_{\rm gauge}\left(t_1,t_0\right)  
        \label{Shapiro_15PN} 
\end{eqnarray}

\noindent 
up to terms of the order ${\cal O}(c^{-4})$, where  
\begin{eqnarray}
	\Delta \tau_{\rm 1PN}^{M_L}\left(t_1,t_0\right) &=& -\,\frac{1}{c}\,\ve{\sigma} \cdot \Delta \ve{x}_{\rm 1PN}^{M_L}\left(t_1,t_0\right),
	\label{Shapiro_Mass_Multipole_definition}
	\\
	\Delta \tau_{\rm 1.5PN}^{S_L}\left(t_1,t_0\right) &=& -\,\frac{1}{c}\,\ve{\sigma} \cdot \Delta \ve{x}_{\rm 1.5PN}^{S_L}\left(t_1,t_0\right),
	\label{Shapiro_Spin_Multipole_definition}
        \\
	\Delta \tau_{\rm gauge}\left(t_1,t_0\right) &=& -\,\frac{1}{c}\,\ve{\sigma} \cdot \Delta \ve{x}_{\rm gauge} \left(t_1,t_0\right).
	\label{Shapiro_Gauge_Term_definition}
	\end{eqnarray}

\noindent
The explicit expressions for the mass-multipole (gravitoelectric) term (\ref{Shapiro_Mass_Multipole_definition}) reads \cite{Kopeikin1997}

\begin{eqnarray}
        \Delta \tau_{\rm 1PN}^{M_L}\left(t_1,t_0\right) &=&
	+ \frac{2 G}{c^3} \frac{\left(-1\right)^l}{l!} \hat{M}_L
\nonumber\\ 
	&& \hspace{-1.9cm} \times \left(\widehat{\partial}_{L} \,\ln \left(r_{\rm N} + c \tau\right)\bigg|_{\tau = \tau_1} 
	- \widehat{\partial}_{L} \,\ln \left(r_{\rm N} + c \tau\right)\bigg|_{\tau = \tau_0}\right), 
	\nonumber\\ 
	\label{Shapiro_Mass_Multipole} 
\end{eqnarray}

\noindent
and the spin-multipole (gravitomagnetic) term (\ref{Shapiro_Spin_Multipole_definition}) reads \cite{Kopeikin1997} (see also footnote $3$ in \cite{Ciufolini}) 
\begin{eqnarray}
	\Delta \tau_{\rm 1.5PN}^{S_L}\left(t_1,t_0\right) &=& + \frac{4 G}{c^4} \frac{\left(-1\right)^l\,l}{\left(l+1\right)!}
	\,\epsilon_{abc}\,\sigma^c\,\hat{S}_{bL-1}
	\nonumber\\ 
	&& \hspace{-2.75cm} \times\!\left(\!\widehat{\partial}_{a L-1} \ln \left(r_{\rm N} + c \tau\right)\bigg|_{\tau = \tau_1} 
	- \widehat{\partial}_{a L- 1} \ln \left(r_{\rm N} + c \tau\right)\bigg|_{\tau = \tau_0}\right). 
	\nonumber\\ 
\label{Shapiro_Spin_Multipole}
\end{eqnarray}

\noindent 
The gauge term (\ref{Shapiro_Gauge_Term_definition}) is determined in Appendix~\ref{Appendix1}, where it is shown that this term vanishes at minus and plus infinity, 
\begin{eqnarray}
        \lim_{\tau = \tau_0 \rightarrow - \infty \atop \tau = \tau_1 \rightarrow + \infty} \Delta \tau_{\rm gauge}\left(t_1,t_0\right)  = 0\;.
        \label{second_integration_gauge}
\end{eqnarray}

\noindent
That means, the gauge terms have no impact on the Shapiro time delay when source and observer are located at spatial infinity, where the
space-time is the flat Minkowski space.
Thus, relation (\ref{second_integration_gauge}) is just a specific example of the general fact, that observables are independent of the chosen gauge.
A similar conclusion is valid for the total light deflection, that is the bending of light in case the source and observer are
located at spatial infinity. In \cite{Zschocke_Total_Light_Deflection_Proceeding} it has been shown that the gauge terms have no
impact on the total light deflection, which is also an observable.

Thus, Eqs.~(\ref{Shapiro_Mass_Multipole}) and (\ref{Shapiro_Spin_Multipole}) represent the effect of time delay and were also given in the textbook \cite{Book_Soffel_Han}. 
In (\ref{Shapiro_Mass_Multipole}) and (\ref{Shapiro_Spin_Multipole}) the differentiations have to be performed. 
Afterwards one has to substitute the unperturbed light ray (\ref{Unperturbed_Lightray_2}) and its absolute value (\ref{Unperturbed_Lightray_Absolute_2}) by the standard 
expressions as given by (\ref{Unperturbed_Lightray_1}) and (\ref{Unperturbed_Lightray_Absolute_1}). 
At the very end of these differentiations, the sub-labels in (\ref{Shapiro_Mass_Multipole}) and (\ref{Shapiro_Spin_Multipole}) are replaced by 
\begin{eqnarray}
	c \tau_0 &=& \ve{\sigma} \cdot \ve{x}_{\rm N}\left(t_0\right),
        \label{sublabel_tau_0}
        \\
	c \tau_1 &=& \ve{\sigma} \cdot \ve{x}_{\rm N}\left(t_1\right), 
        \label{sublabel_tau_1}
\end{eqnarray}

\noindent 
and $P^{ij} \xi_j$ by the spatial components of the impact vector $d^i_{\sigma}$. 
Further details about this approach can be found in \cite{Kopeikin1997,Book_Soffel_Han,Zschocke_Time_Delay_2PN}. 

The time delay in 1.5PN approximation, given by Eq.~(\ref{Shapiro_15PN}), is valid for finite distances of source and observer from the gravitating body. 
According to (\ref{geodesic_equation_15}), the spatial coordinates of source, $\ve{x}_0$, and observer, $\ve{x}_1$, are related to the spatial coordinates 
of the unperturbed light signal at time of emission, $t_0$ and time of observation, $t_1$, as follows, 
\begin{eqnarray}
	\ve{x}_0 &=& \ve{x}_{\rm N}\left(t_0\right),
	\label{source_x0}
	\\
	\ve{x}_1 &=& \ve{x}_{\rm N}\left(t_1\right) + {\cal O}\left(c^{-2}\right).
	\label{observer_x1}
\end{eqnarray}

\noindent
Therefore, after performing the differentiations, one may replace these awkward terms $\ve{x}_{\rm N}\left(t_0\right)$ and $\ve{x}_{\rm N}\left(t_1\right)$ 
in the arguments of (\ref{Shapiro_Mass_Multipole}) and (\ref{Shapiro_Spin_Multipole}) just by the exact positions of source and observer. 

Like in case of total light deflection \cite{Kopeikin1997,Zschocke_Total_Light_Deflection_15PN}, we will determine the total effect of time delay (\ref{Shapiro_Mass_Multipole}) 
and (\ref{Shapiro_Spin_Multipole}). That means, we will consider astrometric configurations, where both the source and the observer are located at spatial infinity from the 
gravitating body. In particular, we will determine the time delay for asymptotic configurations of source and observer, where the limits are 
\begin{eqnarray}
        \ve{\sigma} \cdot \ve{x}_0 &\rightarrow& - \infty\;,
        \label{limit_tau_0}
        \\
        \ve{\sigma} \cdot \ve{x}_1 &\rightarrow& + \infty\;.
        \label{limit_tau_1}
\end{eqnarray}

\begin{figure}[!ht]
\includegraphics[scale=0.105]{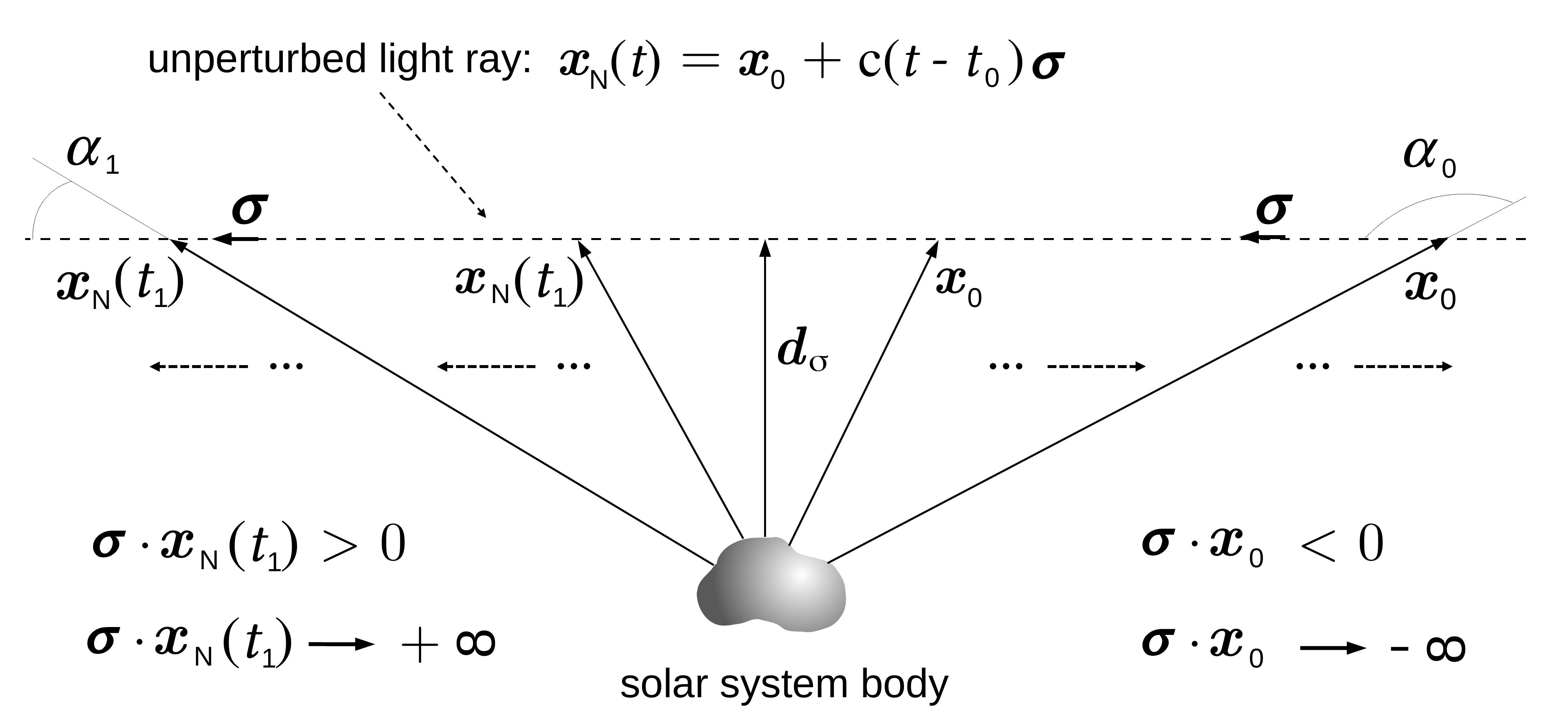}
\caption{This diagram elucidates the limits in (\ref{tau_0}) and (\ref{tau_1}). The angles are $\alpha_0 = \delta\left(\ve{\sigma},\ve{x}_0\right)$ and 
$\alpha_1 = \delta\left(\ve{\sigma}, \ve{x}_{\rm N}\left(t_1\right)\right)$. The dashed arrows to the right and left show the directions to which the spatial positions 
of source, $\ve{x}_0 = \ve{x}_{\rm N}\left(t_0\right)$, and the spatial position of the unperturbed light signal $\ve{x}_{\rm N}\left(t_1\right)$ at time of observation, are shifted 
up to infinity. The limits in (\ref{tau_0}) and (\ref{tau_1}) implicitly assume that the impact vector of the unperturbed light ray remains constant: 
$\ve{d}_{\sigma} = {\rm const}$.}
\label{Diagram2}
\end{figure}

\noindent 
Roughly to speak, these conditions represent configurations, where the massive body is located somewhere \glqq between\grqq{} source and observer, as shown in Fig.~\ref{Diagram2}. 
In view of (\ref{sublabel_tau_0}) - (\ref{limit_tau_1}), these asymptotic limits (\ref{limit_tau_0}) and (\ref{limit_tau_1}) in terms of the arguments of the time delay 
in (\ref{Shapiro_Mass_Multipole}) and (\ref{Shapiro_Spin_Multipole}) reads 
\begin{eqnarray}
c \tau_0 \rightarrow - \infty\,, 
\label{tau_0}
\\
c \tau_1 \rightarrow + \infty\,.
\label{tau_1}
\end{eqnarray}

\noindent 
The time delay (\ref{Shapiro_Mass_Multipole}) and (\ref{Shapiro_Spin_Multipole}) depends on the impact vector of the unperturbed light ray, which is constant for a given light ray. 
Therefore, these limits in (\ref{tau_0}) and (\ref{tau_1}) have to be taken along the unperturbed light trajectory with constant impact vector, as displayed by Fig.~\ref{Diagram2}. 
In particular, one finds the following limits, 
\begin{eqnarray}
	\lim_{\tau_0 \rightarrow - \infty}\,\frac{\ve{\sigma} \cdot \ve{x}_{\rm N}\left(t_0\right)}{r_{\rm N}\left(t_0\right)} &=& - 1\,,
\label{limit_vector_0}
\\
	\lim_{\tau_1 \rightarrow + \infty}\,\frac{\ve{\sigma} \cdot \ve{x}_{\rm N}\left(t_1\right)}{r_{\rm N}\left(t_1\right)} &=& + 1\,,
\label{limit_vector_1}
\end{eqnarray}

\noindent 
where $r_{\rm N}\left(t_0\right) = |\ve{x}_{\rm N}\left(t_0\right)|$ and $r_{\rm N}\left(t_1\right) = |\ve{x}_{\rm N}\left(t_1\right)|$. Thus, the angles are  
$\delta(\ve{\sigma}, \ve{x}_0) \rightarrow \pi$ and $\delta(\ve{\sigma}, \ve{x}_{\rm N}\left(t_1\right)) \rightarrow 0$ in these limits, a result which is 
elucidated by the graphical representation in Fig.~\ref{Diagram2}. 

\noindent 
The mass-monopole term ($l=0$ in (\ref{Shapiro_Mass_Multipole})) has already been given by Eq.~(\ref{Shapiro_1PN}), which is logarithmically divergent. 
Therefore, we consider the asymptotic limit only for mass-multipole terms with $l \ge 1$ and for spin-multipole terms with $l \ge 1$. We use the following notation 
for these asymptotic limits, 
\begin{eqnarray}
	\Delta \tau_{\rm 1PN}^{M_L} &=& \lim_{\tau_0 \rightarrow - \infty \atop \tau_1 \rightarrow + \infty} \Delta \tau_{\rm 1PN}^{M_L}\left(t_1,t_0\right), 
        \label{Shapiro_Mass_Multipole_asymptotic}
	\\
	\Delta \tau_{\rm 1.5PN}^{S_L} &=& \lim_{\tau_0 \rightarrow - \infty \atop \tau_1 \rightarrow + \infty} \Delta \tau_{\rm 1.5PN}^{S_L}\left(t_1,t_0\right), 
	\label{Shapiro_Spin_Multipole_asymptotic}
\end{eqnarray}

\noindent 
where $\Delta c\tau_{\rm 1PN}^{M_L}\left(t_1,t_0\right)$ and $\Delta c\tau_{\rm 1.5PN}^{S_L}\left(t_1,t_0\right)$ were given by Eqs.~(\ref{Shapiro_Mass_Multipole}) and 
(\ref{Shapiro_Spin_Multipole}). In Appendix~\ref{Appendix2} the following results for these limits are shown  
\begin{eqnarray}
	\widehat{\partial}_{L} \,\ln \left(r_{\rm N} + c \tau\right)\bigg|_{\tau = \tau_0 \rightarrow - \infty} &=& 2\,\widehat{\partial}_{L} \,\ln \left| \ve{\xi}\right|, 
        \label{Shapiro_minus_infinity}
	\\
	\widehat{\partial}_{L} \,\ln \left(r_{\rm N} + c \tau\right)\bigg|_{\tau = \tau_1 \rightarrow + \infty} &=& 0\;, 
        \label{Shapiro_plus_infinity}
\end{eqnarray}

\noindent
which are valid for $l \ge 1$. 
These limits in (\ref{Shapiro_minus_infinity}) and (\ref{Shapiro_plus_infinity}) can also nicely be verified for the first few orders in $l$ just by explicit computation. 
Let us notice, that on the left-hand side of these relations one has, first of all, to perform the differentiations by using 
the differential operator in (\ref{Transformation_Derivative_3}) and afterwards one has to calculate the limits. Then, on the right-hand side of these relations only 
the term with $p=0$ in the differential operator (\ref{Transformation_Derivative_3}) contributes, which is given by (\ref{Appendix2_25}) in Appendix~\ref{Appendix2}; 
see also comment in the text below Eq.~(\ref{Appendix_Shapiro_minus_infinity}). 

By inserting (\ref{Shapiro_minus_infinity}) and (\ref{Shapiro_plus_infinity}) into (\ref{Shapiro_Mass_Multipole}) and (\ref{Shapiro_Spin_Multipole}) one obtains for the 
time delay in the asymptotic limit where source and observer are located at spatial infinity, the following expressions:
\begin{eqnarray}
\Delta \tau_{\rm 1PN}^{M_L} &=& - \frac{4 G}{c^3} \frac{\left(-1\right)^l}{l!} \hat{M}_L
        \;\widehat{\partial}_{L} \,\ln\left| \ve{\xi}\right|,  
        \label{Total_Shapiro_Mass_Multipole}
	\\
        \Delta \tau_{\rm 1.5PN}^{S_L} &=& - \frac{8 G}{c^4} \frac{\left(-1\right)^l\,l}{\left(l+1\right)!}
        \,\epsilon_{abc}\,\sigma^c\,\hat{S}_{bL-1}\; \widehat{\partial}_{aL-1} \,\ln\left| \ve{\xi}\right|,
	\nonumber\\ 
\label{Total_Shapiro_Spin_Multipole}	
\end{eqnarray}

\noindent
which are valid for $l \ge 1$. In order to determine the time delay by means of the expressions (\ref{Total_Shapiro_Mass_Multipole}) and (\ref{Total_Shapiro_Spin_Multipole}), 
one has to calculate the term 
\begin{eqnarray}
	\widehat{\partial}_{L} \,\ln\left| \ve{\xi}\right| &=& {\rm STF}_{i_1 \dots i_l}\;P_{i_{1}}^{j_{1}}\;...\;P_{i_l}^{j_l} 
        \,\frac{\partial}{\partial \xi^{j_{1}}}\;...\;\frac{\partial}{\partial \xi^{j_{l}}}\;\ln\left| \ve{\xi}\right|.  
	\nonumber\\ 
\end{eqnarray}

\noindent 
In our investigation \cite{Zschocke_Total_Light_Deflection_15PN} it has been shown that this term is given by the following expression 
\begin{eqnarray}
        \widehat{\partial}_{L}\,\ln \left|\ve{\xi}\right| &=& \frac{\left(-1\right)^{l+1}}{\left| \ve{\xi}\right|^l}\;{\rm STF}_{i_1 \dots i_l}\;  
        \sum\limits_{n=0}^{[l/2]} G_n^l\;P_{i_1 i_2}\, \dots \, P_{i_{2 n - 1} i_{2 n}} 
	\nonumber\\ 
	&& \times \frac{\xi_{i_{2 n + 1}}\,\dots\,\xi_{i_{l}}}{\left|\ve{\xi}\right|^{l-2n}}\;,  
        \label{Relation_C}
\end{eqnarray}

\noindent
which is valid for any natural number $l \ge 1$, and the coefficients are given by \cite{Zschocke_Total_Light_Deflection_15PN}  
\begin{eqnarray}
        G^l_n &=& \left(-1\right)^{n}\,2^{l - 2 n - 1}\,\frac{l!}{n!}\,\frac{\left(l - n - 1\right)!}{\left(l - 2 n\right)!}\;.  
        \label{Relation_D}
\end{eqnarray}

\noindent 
Inserting (\ref{Relation_C}) into (\ref{Total_Shapiro_Mass_Multipole}) and (\ref{Total_Shapiro_Spin_Multipole}) completes the calculation of the 
time delay in the gravitational field of a massive body with full (time-independent) mass-multipole and spin-multipole structure, if the source and 
the observer are at infinite distance from the massive body.

\section{Relation between time delay and total light deflection}\label{Relation}

The tangent vector of the light trajectory at minus infinity, $\ve{\sigma}$, has been defined by Eq.~(\ref{Initial_A}), 
and the tangent vector of the light trajectory at plus infinity, $\ve{\nu}$, is defined by 
\begin{eqnarray}
        \ve{\nu} &=& \frac{\dot{\ve{x}}\left(t\right)}{c}\,\bigg|_{t = + \infty} \quad {\rm with} \quad
        \ve{\nu} \cdot \ve{\nu} = 1\;. 
        \label{vector_nu_5}
\end{eqnarray}

\noindent 
A graphical representation of the three-vector in (\ref{vector_nu_5}) is given in Fig.~\ref{Diagram1}. 
The angle of total light deflection is defined as angle $\delta\left(\ve{\sigma} , \ve{\nu}\right)$ between these tangent vectors, 
\begin{eqnarray}
	\delta\left(\ve{\sigma} , \ve{\nu}\right) &=& \arcsin \left| \ve{\sigma} \times \ve{\nu}\right|. 
	\label{total_light_deflection_angle}
\end{eqnarray}

\noindent 
The tangent vector $\ve{\nu}$ can be expanded in terms of mass-multipoles $\hat{M}_L$ and spin-multipoles $\hat{S}_L$ \cite{Klioner1991,Zschocke_Total_Light_Deflection_15PN}  
\begin{eqnarray}
        \ve{\nu} &=& \ve{\sigma} + \sum\limits_{l=0}^{\infty} \ve{\nu}^{M_L}_{\rm 1PN} + \sum\limits_{l=1}^{\infty} \ve{\nu}^{S_L}_{\rm 1.5PN} + {\cal O}\left(c^{-4}\right),
        \label{vector_nu_10}
\end{eqnarray}

\noindent
where the individual terms are given by Eqs.~(48) and (49) in \cite{Zschocke_Total_Light_Deflection_15PN}. By inserting (\ref{vector_nu_10}) into (\ref{total_light_deflection_angle}) 
one finds that the angle of total light deflection is also expanded in terms of mass-multipoles and spin-multipoles, 
\begin{eqnarray}
        \delta\left(\ve{\sigma} , \ve{\nu}\right) &=& \sum\limits_{l=0}^{\infty}\delta\left(\ve{\sigma} , \ve{\nu}^{M_L}_{\rm 1PN}\right) 
	+ \sum\limits_{l=1}^{\infty} \delta\left(\ve{\sigma} , \ve{\nu}^{S_L}_{\rm 1.5PN}\right) + {\cal O}\left(c^{-4}\right).
	\nonumber\\ 
        \label{total_light_deflection_angle_10}
\end{eqnarray}

\noindent
The individual terms are given by \cite{Kopeikin1997,Zschocke_Total_Light_Deflection_15PN}  
\begin{eqnarray}
        \delta\left(\ve{\sigma}, \ve{\nu}_{\rm 1PN}^{M_L}\right) &=& 
        - \frac{4 G}{c^2}\,\frac{1}{\left|\ve{\xi}\right|} \frac{\left(-1\right)^l}{\left(l-1\right)!} \,
        \hat{M}_{L} \, \widehat{\partial}_{L}\;\ln \left|\ve{\xi}\right| ,
        \label{light_deflection_mass}
        \\
        \delta\left(\!\ve{\sigma}, \ve{\nu}_{\rm 1.5PN}^{S_L}\!\right) &=& 
        - \frac{8 G}{c^3} \frac{1}{\left|\ve{\xi}\right|}\,\frac{\left(-1\right)^l l^2}{\left(l+1\right)!}\,
        \epsilon_{abc}\,\sigma^c\,\hat{S}_{b L-1} \widehat{\partial}_{a L-1} \ln \left|\ve{\xi}\right|,
	\nonumber\\ 
        \label{light_deflection_spin}
\end{eqnarray}

\noindent
which are valid for $l \ge 1$. 
By comparing (\ref{light_deflection_mass}) and (\ref{light_deflection_spin}) with Eqs.~(\ref{Total_Shapiro_Mass_Multipole}) and (\ref{Total_Shapiro_Spin_Multipole}) 
one recovers the following remarkable relations between the multipole terms of the total effect of time delay and the multipole terms of the angle of total light deflection:  
\begin{eqnarray}
	\Delta \tau_{\rm 1PN}^{M_L} &=& \frac{1}{l}\,\frac{\left|\ve{\xi}\right|}{c}\,\delta\left(\ve{\sigma} , \ve{\nu}^{M_L}_{\rm 1PN}\right),  
        \label{Relation_Shapiro_Lightdeflection_M}
        \\
	\Delta \tau_{\rm 1.5PN}^{S_L} &=& \frac{1}{l}\,\frac{\left|\ve{\xi}\right|}{c}\,\delta\left(\ve{\sigma} , \ve{\nu}^{S_L}_{\rm 1.5PN}\right),
        \label{Relation_Shapiro_Lightdeflection_S}
\end{eqnarray}

\noindent
which are valid for multipoles of order $l \ge 1$ and $|\ve{\xi}| = d_{\sigma}$ is the impact parameter of the unperturbed light ray (\ref{absolute_value_impact_vector_d_sigma}). 
These relations (\ref{Relation_Shapiro_Lightdeflection_M}) and (\ref{Relation_Shapiro_Lightdeflection_S}) are strictly valid for light signals which propagate in the gravitational 
fields of a body at rest, having the full set of mass-multipoles and spin-multipoles. That means, the body can be of arbitrary shape, inner structure and can be in arbitrary 
but uniform rotational motions and stationary inner currents.

\section{Time delay in field of an axisymmetric body}\label{Section_Shapiro_2}

\begin{figure}[!ht]
\includegraphics[scale=0.12]{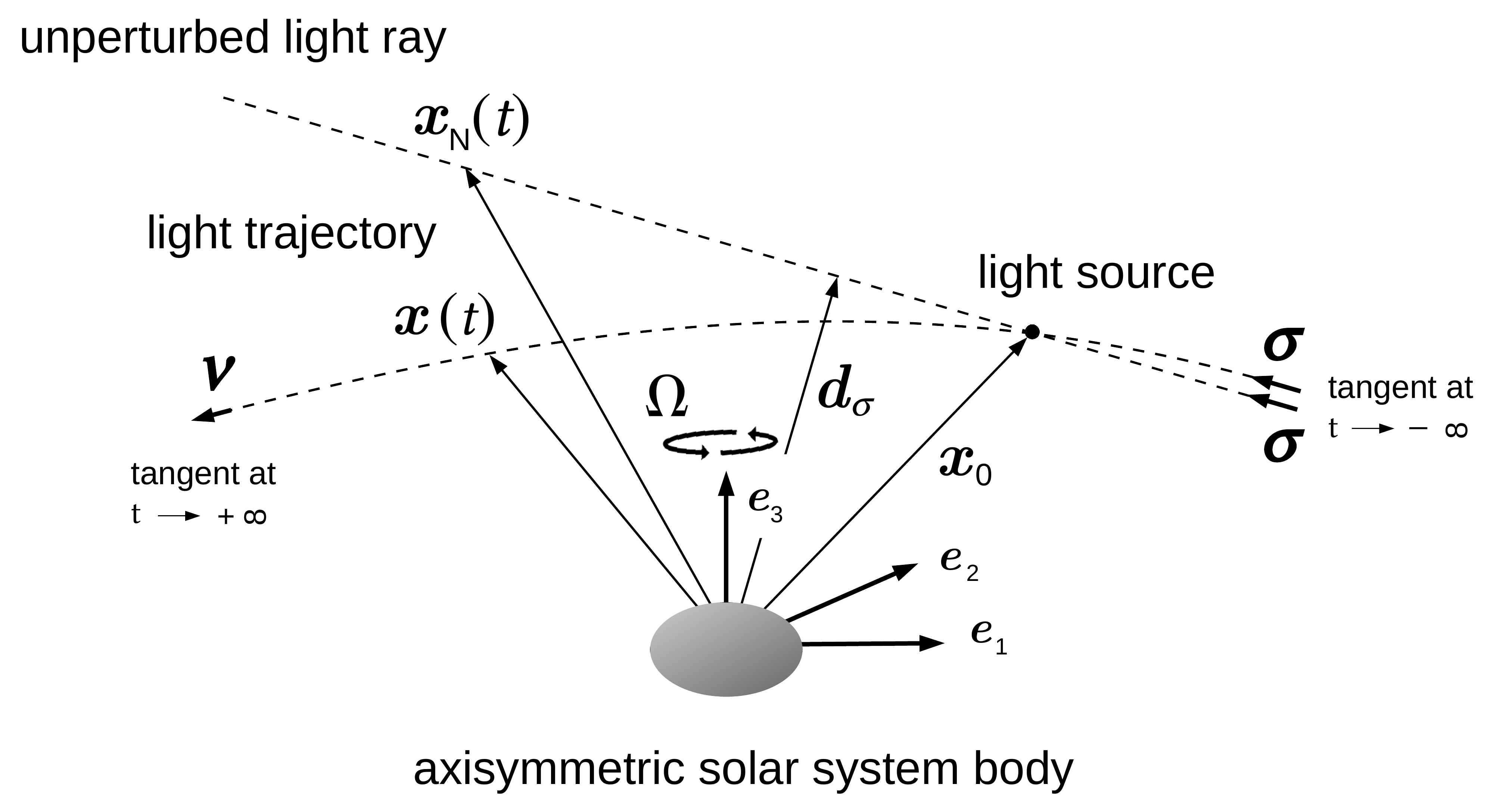}
        \caption{A geometrical representation of the propagation of a light signal through the gravitational field of an axisymmetric Solar System body
	at rest. The axes of inertia are denoted by $\ve{e}_1, \ve{e}_2, \ve{e}_3$. 
	The body is in uniform rotational motion with angular velocity $\Omega$ around the axis of symmetry $\ve{e}_3$. 
	The light signal is emitted by the light source at $\ve{x}_0$ and propagates along the exact light trajectory $\ve{x}\left(t\right)$.
	The unit tangent vectors along the light trajectory at minus and plus infinity are $\ve{\sigma}$ and $\ve{\nu}$.
        The unperturbed light ray $\ve{x}_{\rm N}\left(t\right)$ is given by Eq.~(\ref{Unperturbed_Lightray_1}) and propagates in the direction
        of $\ve{\sigma}$ along a straight line through the position of the light source at $\ve{x}_0$. The impact vector $\ve{d}_{\sigma}$ of the
        unperturbed light ray is given by Eq.~(\ref{impact_vector_d_sigma}).}
\label{Diagram3}
\end{figure}

The largest effect of Shapiro effect is expected from the Sun and the giant planets of the Solar System. In order to determine the Shapiro time delay 
one needs the explicit form for mass-multipoles (\ref{Mass_Multipoles}) and for spin-multipoles (\ref{Spin_Multipoles}). 
For an estimation of the individual terms in (\ref{Total_Shapiro_Mass_Multipole}) and (\ref{Total_Shapiro_Spin_Multipole}), one may approximate the Sun and the giant planets 
by a rigid axisymmetric body with radial dependent mass distribution and in uniform rotational motion with angular velocity $\Omega$ around the symmetry axis of the body, 
which is aligned with the $x^3$-axis of the coordinate system. A graphical representation of this configuration is given by Fig.~\ref{Diagram3}. 
Then, the higher mass-multipoles and spin-multipoles for such a body are given by \cite{Zschocke_Time_Delay_2PN,Zschocke_Total_Light_Deflection_15PN}  
\begin{eqnarray}
\hat{M}_L &=& - M \left(P\right)^l\,J_l\;\delta^3_{<{i_1}} \; \dots \; \delta^3_{{i_l}>}\;, 
\label{M_L_Newtonian}
\\ 
\hat{S}_{a} &=& + \kappa^2\,M\,P^2\,\Omega\,\delta_{3a}\;,
\label{S_1_Newtonian}
\\ 
	\hat{S}_L &=& - M \left(P\right)^{l+1}\,\Omega\,J_{l-1}\,\frac{l+1}{l+4}\;\delta^3_{<{i_1}} \; \dots \; \delta^3_{{i_l}>}\;, 
\label{S_L_Newtonian}
\end{eqnarray}

\noindent 
where (\ref{M_L_Newtonian}) is valid for any natural number of $l \ge 2$, while (\ref{S_L_Newtonian}) is valid for any natural number of $l \ge 3$. 
The mass-dipole term, that is $l = 1$ in (\ref{M_L_Newtonian}) vanishes in case the origin of coordinate system is located at the center of mass of the body (i.e. $J_1 = 0$) 
and will, therefore, not be considered in what follows. 
Here, $M$ is the Newtonian mass of the body, $P$ its equatorial radius, $J_l$ are the actual zonal harmonic coefficients of index $l$, 
$\kappa^2$ is the dimensionless moment of inertia, $\Omega$ is the angular velocity of the rotating body and 
$\delta^3_{<{i_1}} \; \dots \; \delta^3_{{i_l}>} = {\rm STF}_{i_1 \dots i_l}\;\delta_{3 i_1} \dots \delta_{3 i_l}$ 
denotes products of Kronecker symbols which are symmetric and traceless with respect to indices $i_1 \dots i_l$. 
It is noticed that the definition of the angular velocity $\Omega$ has an unambiguous meaning only at linear order around flat space-time.

For reason of completeness, we notice the upper limit of the Shapiro time delay caused by the mass-monopole (\ref{Shapiro_1PN}), which reads \cite{Book_Clifford_Will} 
\begin{eqnarray}
        \left|\Delta \tau_{\rm 1PN}^{M_0}\right|  &\le&
	\frac{2 G M}{c^3}\,\ln \frac{4\,x_0\,x_1}{\left(d_{\sigma}\right)^2} \;. 
	\label{Total_Time_Delay_Mass_Monopole}
\end{eqnarray}

\noindent
Now we consider the upper limits of the higher mass-multipoles and spin-multipoles, that means the upper limits of the absolute values in the asymptotic limits 
(\ref{Shapiro_Mass_Multipole_asymptotic}) and (\ref{Shapiro_Spin_Multipole_asymptotic}). In order to get the total effect of time delay in the gravitational field 
of an axisymmetric body in uniform rotational motion, one has to insert the multipoles (\ref{M_L_Newtonian}) - (\ref{S_L_Newtonian}) into Eqs.~(\ref{Total_Shapiro_Mass_Multipole}) 
and (\ref{Total_Shapiro_Spin_Multipole}) and taking account of relation (\ref{Relation_C}). However, in view of relations (\ref{Relation_Shapiro_Lightdeflection_M}) and 
(\ref{Relation_Shapiro_Lightdeflection_S}), which are valid for bodies of arbitrary shape, inner structure and uniform rotational motions, it is easier to use the results for the 
total light deflection angle which have been determined for axisymmetric bodies in \cite{Zschocke_Total_Light_Deflection_15PN}. 
In particular, in our investigation \cite{Zschocke_Total_Light_Deflection_15PN} 
it has been found that the angle of total light deflection is related to Chebyshev polynomials. This relation has been established for mass-multipoles 
by Eq.~(114) in \cite{Zschocke_Total_Light_Deflection_15PN} and for spin-multipoles by Eq.~(121) in \cite{Zschocke_Total_Light_Deflection_15PN}. 
Hence, one obtains for the time delay of light signals in the field of an axisymmetric body the following expressions in the asymptotic limit, 
\begin{eqnarray}
        \Delta \tau_{\rm 1PN}^{M_L}  &=&  
	- \frac{4 G M}{c^3} \frac{J_l}{l} \left(\frac{P}{d_{\sigma}}\right)^l
	\nonumber\\ 
	&& \times \left[1 - \left(\ve{\sigma} \cdot \ve{e}_3\right)^2\right]^{[l/2]} T_l\left(x\right),  
        \label{Total_Time_Delay_Mass_Chebyshev}
\end{eqnarray}

\noindent 
for the mass-multipole terms, which is valid for $l \ge 2$. For the effect of time delay caused by the spin-dipole one obtains, 
by means of Eq.~(120) in \cite{Zschocke_Total_Light_Deflection_15PN}, the following expression in the asymptotic limit, 
\begin{eqnarray}
        \Delta \tau_{\rm 1.5PN}^{S_1} &=&
        - \frac{4 G M}{c^4}\,P\,\Omega\,\kappa^2\,J_0 \left(\frac{P}{d_{\sigma}}\right)\,
        \frac{\left(\ve{\sigma} \times \ve{d}_{\sigma}\right) \cdot \ve{e}_3}{d_{\sigma}}\;,
        \nonumber\\ 
        \label{Total_Time_Delay_Spin_Dipole}
\end{eqnarray}

\noindent
with $J_0 = -1$, and for the effect of time delay caused by higher spin-multipoles,
\begin{eqnarray}
	\Delta \tau_{\rm 1.5PN}^{S_L} &=& 
        - \frac{8 G M}{c^4}\,P\,\Omega\,J_{l-1} \left(\frac{P}{d_{\sigma}}\right)^l \, 
        \frac{\left(\ve{\sigma} \times \ve{d}_{\sigma}\right) \cdot \ve{e}_3}{d_{\sigma}} 
        \frac{1}{l+4} 
	\nonumber\\ 
	&& \times \left[1 - \left(\ve{\sigma} \cdot \ve{e}_3\right)^2\right]^{[l/2]}\;U_{l-1}\left(x\right), 
        \label{Total_Time_Delay_Spin_Chebyshev}
\end{eqnarray}

\noindent
which is valid for $l \ge 3$. The variable in (\ref{Total_Time_Delay_Mass_Chebyshev}) and (\ref{Total_Time_Delay_Spin_Chebyshev}) reads \cite{Zschocke_Total_Light_Deflection_15PN}  
\begin{eqnarray}
        x &=& \left(1 - \left(\ve{\sigma} \cdot \ve{e}_3\right)^2\right)^{-1/2}\;\left(\frac{\ve{d}_{\sigma} \cdot \ve{e}_3}{d_{\sigma}}\right),  
        \label{Variable_x_Chebyshev_A}
\end{eqnarray}

\noindent
which is a real number. It has already been shown in \cite{Zschocke_Total_Light_Deflection_15PN} that the interval of the argument is, in fact, given by $- 1 \le x \le + 1\;$.
The power representation of Chebyshev polynomials of first kind reads \cite{Arfken_Weber} 
\begin{eqnarray}
        T_l \left(x\right) &=& \frac{l}{2} \sum \limits_{n=0}^{[l/2]} \frac{\left(-1\right)^n}{n!} \,\frac{\left(l - n - 1\right)!}{\left(l - 2 n\right)!}
        \,\left(2 x\right)^{l - 2 n}\,,
        \label{Chebyshev_Polynomials_1}
\end{eqnarray}

\noindent
where $l \ge 1$.
The power representation of Chebyshev polynomials of second kind reads \cite{Arfken_Weber} 
\begin{eqnarray}
        U_l \left(x\right) &=& \sum \limits_{n=0}^{[l/2]} \frac{\left(-1\right)^n}{n!} \,\frac{\left(l - n\right)!}{\left(l - 2 n\right)!}
        \,\left(2 x\right)^{l - 2 n}\;, 
        \label{Chebyshev_Polynomials_2}
\end{eqnarray}

\noindent
where $l \ge 0$. 
The remarkable feature that the total effect of time delay is given in terms of Chebyshev polynomials allows for a straightforward determination of the upper limit of the
total effect of time delay, because the upper limits of Chebyshev polynomials are given by
\begin{eqnarray}
        \left|T_l\right| \le 1 \quad {\rm and} \quad \left|U_{l-1}\right| \le l \;.
        \label{Introduction_Chebyshev_polynomials_2}
\end{eqnarray}

\noindent
By inserting (\ref{Introduction_Chebyshev_polynomials_2}) into (\ref{Total_Time_Delay_Mass_Chebyshev}) and (\ref{Total_Time_Delay_Spin_Chebyshev}) yields for the 
absolute values of the Shapiro time delay induced by the mass-multipoles and spin-multipoles the following inequalities: 
for the mass-multipole terms ($l \ge 2$) one obtains, 
\begin{eqnarray}
        \left|\Delta \tau_{\rm 1PN}^{M_L}\right| &\le& \frac{4 G M}{c^3}\,\frac{\left|J_l\right|}{l} \left( 1 - \left(\ve{\sigma} \cdot \ve{e}_3\right)^2  \right)^{[l/2]}
	\left(\frac{P}{d_{\sigma}}\right)^l , 
        \label{upper_limit_M_L}
\end{eqnarray}
\noindent
for the spin-dipole term ($l=1$) one finds, 
\begin{eqnarray}
        \left|\Delta \tau_{\rm 1.5PN}^{S_1}\right| &\le& \frac{4 G M}{c^4} P\,\Omega\,\kappa^2 
        \left(1 - \left(\ve{\sigma} \cdot \ve{e}_3\right)^2\right)^{[l/2]} \left(\frac{P}{d_{\sigma}}\right) ,
        \nonumber\\
        \label{upper_limit_S_1}
\end{eqnarray}

\noindent 
and for the spin-multipole terms ($l \ge 3$) one obtains, 
\begin{eqnarray}
        \left|\Delta \tau_{\rm 1.5PN}^{S_L}\right| &\le& \frac{8 G M}{c^4} P\,\Omega \frac{\left|J_{l-1}\right| l}{l+4} 
        \left(1 - \left(\ve{\sigma} \cdot \ve{e}_3\right)^2\right)^{[l/2]}\!\left(\frac{P}{d_{\sigma}}\right)^{l} . 
	\nonumber\\ 
        \label{upper_limit_S_L}
\end{eqnarray}

\noindent
These upper limits (\ref{upper_limit_M_L}) and (\ref{upper_limit_S_L}) represent a criterion to identify those multipoles which contribute significantly 
to the time delay for a given accuracy of time measurements. By taking into account that $|\,1 - \left(\ve{\sigma} \cdot \ve{e}_3\right)^2 | \le 1$ one may derive simpler 
expressions for these upper limits, namely for the mass-multipoles ($l \ge 2$), 
\begin{eqnarray}
        \left|\Delta \tau_{\rm 1PN}^{M_L}\right| &\le& \frac{4 G M}{c^3}\,\frac{\left|J_l\right|}{l} \,\left(\frac{P}{d_{\sigma}}\right)^l ,
        \label{upper_limit_M_L_B}
\end{eqnarray}
\noindent
for the spin-dipole ($l=1$), 
\begin{eqnarray}
        \left|\Delta \tau_{\rm 1.5PN}^{S_1}\right| &\le& \frac{4 G M}{c^4} P\,\Omega\,\kappa^2 \left(\frac{P}{d_{\sigma}}\right) ,
        \label{upper_limit_S_1_B}
\end{eqnarray}

\noindent 
and for the spin-multipoles ($l \ge 3$), 
\begin{eqnarray}
        \left|\Delta \tau_{\rm 1.5PN}^{S_L}\right| &\le& \frac{8 G M}{c^4}\,P\,\Omega\,\frac{l}{l+4}\,\left|J_{l-1}\right|\,
        \left(\frac{P}{d_{\sigma}}\right)^{l} \;.
        \label{upper_limit_S_L_B}
\end{eqnarray}

\noindent
These upper limits have also been presented by Eqs.~(39) and (41) in \cite{Zschocke_Time_Delay_2PN}. However, the 
coefficients in front of Eqs.~(39) and (41) in \cite{Zschocke_Time_Delay_2PN} were only given for the very few first multipoles for ellipsoidal bodies, 
while here these upper limits in (\ref{upper_limit_M_L_B}) and (\ref{upper_limit_S_L_B}) are valid for any multipole order and for the more general case of 
axisymmetric bodies. 

Finally, we notice the upper limits for grazing rays, for the mass-multipole terms ($l \ge 2$) given by 
\begin{eqnarray}
        \left|\Delta \tau_{\rm 1PN}^{M_L}\right| &\le& \frac{4 G M}{c^3}\,\frac{\left|J_l\right|}{l}\,,
        \label{upper_limit_M_L_C}
\end{eqnarray}
\noindent
for the spin-dipole ($l=1$) given by 
\begin{eqnarray}
        \left|\Delta \tau_{\rm 1.5PN}^{S_1}\right| &\le& \frac{4 G M}{c^4} P\,\Omega\,\kappa^2\;,
        \label{upper_limit_S_1_C}
\end{eqnarray}

\noindent 
and for the spin-multipole terms ($l \ge 3$) one obtains
\begin{eqnarray}
        \left|\Delta \tau_{\rm 1.5PN}^{S_L}\right| &\le& \frac{8 G M}{c^4}\,P\,\Omega\,\frac{l}{l+4}\,\left|J_{l-1}\right|\,.
        \label{upper_limit_S_L_C}
\end{eqnarray}

\noindent
Numerical values of these limits (\ref{upper_limit_M_L_C}) - (\ref{upper_limit_S_L_C}) are given by Table~\ref{Table2} for Solar System bodies.

\section{Comparison with the literature}\label{Section_Comparison}

The upper limits of Shapiro time delay for mass-monopole (\ref{Total_Time_Delay_Mass_Monopole}) and spin-dipole (\ref{upper_limit_S_1_C}) are well-known and agree, 
for instance, with Eq.~(40.13) in \cite{MTW} and Eq.~(75) in \cite{Klioner1991}, respectively. 

Upper limits for the effect of time delay caused by higher mass-multipoles and spin-multipoles of the massive body are rare and have, thus far, 
only been presented in our previous investigation in \cite{Zschocke_Time_Delay_2PN}. 
In order to compare the upper limits in (\ref{upper_limit_M_L_B}) and (\ref{upper_limit_S_L_B}) with those upper limits
presented by Eqs.~(39) and (41) in \cite{Zschocke_Time_Delay_2PN}, we rewrite (\ref{upper_limit_M_L_B}) and (\ref{upper_limit_S_L_B}) in the following form:
\begin{eqnarray}
        \left|\Delta \tau_{\rm 1PN}^{M_L}\right| &\le& A_l\,\frac{G M}{c^3}\,\left|J_l\right| \,\left(\frac{P}{d_{\sigma}}\right)^l \; {\rm with} \;
        A_l = \frac{4}{l}\;,
        \label{comparison_upper_limit_M_L}
\end{eqnarray}
\noindent
which is valid for $l \ge 2$, and for the spin-multipole terms one obtains
\begin{eqnarray}
        \left|\Delta \tau_{\rm 1.5PN}^{S_L}\right| &\le& B_l\,\frac{G M}{c^4}\,P\,\Omega\,\left|J_{l-1}\right|\,
        \left(\frac{P}{d_{\sigma}}\right)^{l} \; {\rm with} \;
        B_l = \frac{8\,l}{l+4}\;,
	\nonumber\\ 
        \label{comparison_upper_limit_S_L}
\end{eqnarray}

\noindent
which is valid for $l \ge 3$. These coefficients, for the first few mass-multipoles and spin-multipoles read
\begin{eqnarray}
        A_2 &=& 2\,, \quad A_4 = 1\,,\quad A_6 = \frac{2}{3}\,, \quad A_8 = \frac{1}{2}\,,\quad A_{10} = \frac{2}{5}\,, 
	\nonumber\\
	B_3 &=& \frac{24}{7}\,.
        \label{coefficients_A_B}
\end{eqnarray}

\noindent
In Eqs.~(42) and (43) in \cite{Zschocke_Time_Delay_2PN} the following coefficients were presented:
\begin{eqnarray}
        A_2 &=& \frac{11}{5}\,, \quad A_4 = \frac{7}{6}\,,\quad A_6 = \frac{3}{5}\,, \quad A_8 = \frac{3}{10}\,,\quad A_{10} = \frac{3}{20}\,, 
	\nonumber\\
	B_3 &=& \frac{7}{6}\,.
        \label{coefficients_zschocke}
\end{eqnarray}

\noindent
The coefficients in (\ref{coefficients_zschocke}) have been obtained for the case of finite spatial distances of source and observer from the massive body, while 
the coefficients in (\ref{coefficients_A_B}) are valid for infinite spatial distances of source and observer from the massive body. Since the general case of 
finite distances contains also the specific case of infinite distances, one would expect that the coefficients in (\ref{coefficients_zschocke}) are slightly larger 
than the coefficients in (\ref{coefficients_A_B}). In fact, this is the case for the multipole orders $l=2$ and $l=4$, while the coefficients for $l=6$ in 
(\ref{coefficients_zschocke}) and (\ref{coefficients_A_B}) are almost equal. These facts are also reflected by the numerical values presented 
here in Table II and in Table II in our previous investigation in \cite{Zschocke_Time_Delay_2PN}. On the other side, the upper limits of mass-multipole terms of order $l=8$ and $l=10$ 
and the upper limit of the spin-multipole term of order $l=3$, presented in our previous investigation in \cite{Zschocke_Time_Delay_2PN}, are too small. 
These deviations have carefully been analyzed and were caused by an inaccuracy in the analytical calculations in \cite{Zschocke_Time_Delay_2PN}, 
which were assisted by computer algebra systems because of the involved algebraic structure of the expressions for the Shapiro time delay, especially in case of higher multipoles 
and for finite distances.

\section{Numerical values of time delay in the Solar System}\label{Section_Numerical_Values}

For the numerical values of time delay we take the parameter of Solar System bodies as given by Table~\ref{Table1}.
\begin{table}[h!]
\caption{Numerical parameter for Schwarzschild radius $G M / c^2$, equatorial radius $P$ and actual zonal harmonic coefficients $J_l$ of Sun, Jupiter, and Saturn.
The values for $G M/c^2$ and equatorial $P$ are taken from \cite{Ellipticity}. The value for $J_l$ for the Sun are from \cite{J_n_Sun} and references therein.
The values $J_l$ with $n=2,4,6$ for Jupiter and Saturn are taken from \cite{Book_Zonal_Harmonics},
while $J_l$ with $n=8$ for Jupiter and Saturn come from \cite{Zonal_Harmonics_Jupiter} and \cite{Zonal_Harmonics_Saturn}, respectively.
The angular velocities $\Omega = 2 \pi /T$ (with rotational period $T$) are presented by NASA planetary fact sheets.
The dimensionless moment of inertia $\kappa^2$ is defined by Eq.~(\ref{kappa}) and their values have been determined in \cite{Ellipticity}.} 
\begin{tabular}{| c | c | c | c |}
\hline
&&&\\[-12pt]
Parameter
&\hbox to 20mm{\hfill Sun \hfill}
&\hbox to 20mm{\hfill Jupiter \hfill}
&\hbox to 20mm{\hfill Saturn \hfill}\\[3pt]
\hline
&&&\\[-12pt]
$GM/c^2\,[{\rm m}]$ & $1476.8$ & $1.41$ & $0.42$ \\[3pt]
$P\,[{\rm m}]$ & $696 \times 10^6$ & $71.5 \times 10^6$ & $60.3 \times 10^6$ \\[3pt]
$J_2$ & $1.7 \times 10^{-7}$ & $14.696 \times 10^{-3}$ & $16.291 \times 10^{-3}$ \\[3pt]
$J_4$ & $ 9.8 \times 10^{-7} $ & $ - 0.587 \times 10^{-3}$ & $ - 0.936 \times 10^{-3}$ \\[3pt]
$J_6$ & $ 4 \times 10^{-8} $ & $0.034 \times 10^{-3}$ & $0.086 \times 10^{-3}$ \\[3pt]
$J_8$ & $ - 4 \times 10^{-9} $ & $ - 2.5 \times 10^{-6}$ & $ - 10.0 \times 10^{-6}$ \\[3pt]
$\Omega\,[{\rm sec}^{-1}]$ & $2.865 \times 10^{-6}$ & $1.758 \times 10^{-4}$ & $1.638 \times 10^{-4}$ \\[3pt]
$\kappa^2$ & $0.059$ & $0.254$ & $0.210$ \\[3pt]
\hline
\end{tabular}
\label{Table1}
\end{table}

The parameter $\kappa^2$ in (\ref{S_1_Newtonian}) is defined by \cite{Ellipticity} (see also Eqs.~(B60) - (B62) in \cite{Zschocke_Time_Delay_2PN})
\begin{eqnarray}
         \kappa^2 &=& \frac{I}{M\,P^2}\;, 
        \label{kappa}
\end{eqnarray}

\noindent
where $I$ is the moment of inertia of the real Solar System body under consideration, which is related to the body's angular momentum via $|\ve{S}| = I\,\Omega$. 
For a spherically symmetric body with uniform density $\kappa^2 = 2/5$ \cite{Ellipticity}, 
while for real Solar System bodies $\kappa^2 < 2/5$ because the mass densities are increasing towards the center of the massive bodies.
The values of $\kappa^2$ are given in Table~\ref{Table1} for the Sun and giant planets of the Solar System bodies.

Numerical values of the upper limits in (\ref{upper_limit_M_L_C}) - (\ref{upper_limit_S_L_C}) are presented in Table~\ref{Table2} for the first mass-multipoles
and spin-multipoles in case of grazing rays at the Sun and the giant planets of the Solar System.
\begin{table}[h!]
        \caption{The effect of (one-way) Shapiro time delay caused by the mass-multipole $\Delta \tau^{M_l}_{\rm 1PN}$ and spin-multipole terms $\Delta \tau^{S_l}_{\rm 1.5PN}$
        in the gravitational field of the Sun and giant planets of the Solar System according to the upper limits (\ref{upper_limit_M_L_C}) - (\ref{upper_limit_S_L_C}). 
        The time delay is given in units of pico-seconds: $1\,{\rm ps} = 10^{-12}\,{\rm sec}$. A blank entry means less than $0.001$ pico-seconds. 
        The values are given for grazing rays (impact parameter $d_{\sigma}$ equals body's equatorial radius $P$).
	The numerical values should be compared with the assumed goal accuracy of $0.001\,{\rm ps}$ in time-delay measurements.} 
\footnotesize
\begin{tabular}{@{}cccccccc}
\hline
&&&&&&\\[-7pt]
        Object & $\Delta \tau^{M_2}_{\rm 1PN}$ & $\Delta \tau^{M_4}_{\rm 1PN}$ & $\Delta \tau^{M_6}_{\rm 1PN}$ & $\Delta \tau^{M_8}_{\rm 1PN}$ & $\Delta \tau^{S_1}_{\rm 1.5PN}$  & $\Delta \tau^{S_3}_{\rm 1.5PN}$\\
&&&&&&\\[-7pt]
\hline
        Sun & $1.68$ & $ 4.83 $ & $ 0.13 $ & $ 0.01 $ & $ 7.73 $ & $ - $ \\
        Jupiter & $138.24$ & $ 2.76 $ & $ 0.11 $ & $ 0.01 $ & $ 0.20 $ & $ 0.010 $ \\
        Saturn & $45.65$ & $ 1.31 $ & $ 0.08 $ & $ 0.01 $ & $ 0.04 $ & $ 0.003 $ \\
\hline
\end{tabular}\\
\label{Table2}
\end{table}
\normalsize

\newpage

\section{Summary}\label{Section_Summary}

In this investigation, the time delay in the gravitational field of a body at rest with full multipole structure has been considered.
In particular, the impact of higher mass-multipoles and spin-multipoles on time delay has been determined. Two main results were found: 
\begin{enumerate}  
\item[(1)] If the source and the observer are located at spatial infinity from the massive body, then the individual multipole terms of time delay are related to the 
individual multipole terms of total light deflection. These relations are given by (\ref{Relation_Shapiro_Lightdeflection_M}) and (\ref{Relation_Shapiro_Lightdeflection_S}), 
which are valid for multipoles of order $l \ge 1$. 
\item[(2)] In case of an axisymmetric massive body, these individual multipole terms of time delay can be expressed in terms of Chebyshev polynomials: mass-multipole terms 
of time delay are related to Chebyshev polynomials of first kind (\ref{Total_Time_Delay_Mass_Chebyshev}) and spin-multipole terms of time delay are related to Chebyshev polynomials 
of second kind (\ref{Total_Time_Delay_Spin_Chebyshev}). 
\end{enumerate} 
These remarkable facts allow one to determine strict upper limits for the absolute value of the multipole terms in this asymptotic configuration, where the source and the observer 
are located at spatial infinity from the massive body. These strict upper limits are given by Eqs.~(\ref{upper_limit_M_L_C}) and (\ref{upper_limit_S_L_C}), which represent a 
criterion to identify those multipoles which contribute significantly to the time delay for a given accuracy of time measurements. The coincidence between the total effect of 
time delay and total light deflection for higher multipoles, as demonstrated by relations (\ref{Relation_Shapiro_Lightdeflection_M}) and (\ref{Relation_Shapiro_Lightdeflection_S}), 
must have a deep reason, which represents a problem for subsequent investigations. 

Numerical values of the Shapiro time delay, based on these upper limits, have been calculated and presented in Table~\ref{Table2}. These numerical results show that the impact 
of the first mass-multipoles with $l \le 8$ and the first spin-multipoles with $l \le  3$ on the effect of time-delay are relevant for an accuracy on the femto-second scale of 
accuracy in time measurements, and might in principle be detected with present-day atomic clocks or, at least, with the next generation of atomic clocks. 

The more realistic astrometric configurations, where source and observer are located at finite spatial distances from the massive body, will be considered in a subsequent investigation. 
It is, however, to expect that the upper limits of higher multipoles on the effect of time delay for such scenarios will not much be different from the asymptotic limit which has been   
considered in this investigation.

\section{Acknowledgment}

This work was funded by the German Research Foundation (Deutsche Forschungsgemeinschaft DFG) under grant number 447922800. Sincere gratitude is expressed to
Prof. Sergei A. Klioner, Prof. Michael H. Soffel, Prof. Ralf Sch\"utzhold, Prof. William G. Unruh, Priv.-Doz. G\"unter Plunien,
Dr. Alexey Butkevich, Dipl-Inf. Robin Geyer, Prof. Burkhard K\"ampfer, and Prof. Laszlo Csernai for inspiring discussions about astrometry and general theory of relativity.

\appendix

\section{Notation}\label{Appendix0}

The following notation is in use:
\begin{itemize}
\item Newtonian constant of gravitation: $G$.
\item vacuum speed of light in flat space-time: $c$.
\item Newtonian mass of the body: $M$.
\item Equatorial radius of the body: $P$.
\item Angular velocity of the body: $\Omega$.
\item Zonal harmonic coefficients of the body: $J_l$.
\item $\eta_{\alpha\beta} = \eta^{\alpha\beta}$ is the metric tensor of flat space-time.
\item $g^{\alpha\beta}$ and $g_{\alpha\beta}$ are the contravariant and covariant components of the metric.
\item $g = {\rm det}\left(g_{\mu\nu}\right)$ is the determinant of the metric.  
\item $n! = n \left(n-1\right)\left(n-2\right)\cdot\cdot\cdot 2 \cdot 1$ is the factorial; by definition: $0! = 1$.
\item Lower case Greek indices take values 0,1,2,3.
\item The contravariant components of four-vectors: $a^{\mu} = \left(a^0,a^1,a^2,a^3\right)$.
\item Lower case Latin indices take values 1,2,3.
\item The three-dimensional coordinate quantities (three-vectors) referred to
the spatial axes of the reference system are in boldface: $\ve{a}$.
\item The contravariant components of three-vectors: $a^{i} = \left(a^1,a^2,a^3\right)$.
\item The absolute value of a three-vector:
$a = |\ve{a}| = \sqrt{a^1\,a^1+a^2\,a^2+a^3\,a^3}$.
\item The scalar product of two three-vectors:
$\ve{a}\,\cdot\,\ve{b}=\delta_{ij}\,a^i\,b^j=a^i\,b^i$ with Kronecker delta $\delta_{ij}$.
\item The angle between two three-vectors $\ve{a}$ and $\ve{b}$ is designated as $\delta\left(\ve{a},\ve{b}\right)$. 
\end{itemize}


\section{Proof of Eq.~(\ref{second_integration_gauge})}\label{Appendix1}

In this appendix we will demonstrate the limit (\ref{second_integration_gauge}).
The gauge terms in the geodesic equation (\ref{geodesic_equation_10}) are given by Eq.~(\ref{geodesic_equation_gauge}), which consists of two pieces, 
\begin{eqnarray}
	\ddot{\ve{x}}_{\rm gauge} &=& \ddot{\ve{x}}_{\rm g1} + \ddot{\ve{x}}_{\rm g2}\;.
	\label{Appendix_C_1}
\end{eqnarray}

\noindent 
Their spatial components are given by
\begin{eqnarray}
	\frac{\ddot{x}_{\rm g1}^i \left(t\right)}{c^2} &=& + \partial_j\,w^0_{\,,\,k} \,\sigma^i \sigma^j \sigma^k \;,
        \label{Term1}
	\\
	\frac{\ddot{x}_{\rm g2}^i \left(t\right)}{c^2} &=& - \partial_j\,w^i_{\,,\,k} \,\sigma^j \sigma^k\;,
        \label{Term2}
\end{eqnarray}

\noindent
where the gauge vectors are given by Eqs.~(\ref{gaugefunction1}) and (\ref{gaugefunction2}). Let us consider the first term (\ref{Term1}). Using
$\left(r^{-1}\right)_{\,,\,jk} = 3 x_j x_k / r^5 - \delta_{jk} / r^3$, one obtains
\begin{eqnarray}
	\frac{\ddot{\ve{x}}_{\rm g1} \left(t\right)}{c^2} &=& 
	+ \frac{8 G}{c^3} \sum\limits_{l=0}^{\infty} \frac{\left(-1\right)^l}{l!} \hat{\partial}_L\,\frac{\hat{W}_L}{\left(r_{\rm N}\right)^3}\,\ve{\sigma}
	\nonumber\\ 
	&& - \frac{12 G}{c^3} \sum\limits_{l=0}^{\infty} \frac{\left(-1\right)^l}{l!} \hat{\partial}_L\,\frac{\hat{W}_L}{\left(r_{\rm N}\right)^5}\,\left(d_{\sigma}\right)^2\,\ve{\sigma}\;.
\label{appendix_15}
\end{eqnarray}

\noindent
In (\ref{appendix_15}) the replacements $\ve{x} \rightarrow \ve{x}_{\rm N}$ and $r \rightarrow r_{\rm N} = |\ve{x}_{\rm N}|$ have been performed 
(cf. text below Eq.~(\ref{geodesic_equation_gauge})), where the unperturbed light ray $\ve{x}_{\rm N}$ is given by Eq.~(\ref{Unperturbed_Lightray_1}) 
and its absolute value by Eq.~(\ref{Unperturbed_Lightray_Absolute_1}). 
In addition, the relation $\left(\ve{\sigma} \cdot \ve{x}_{\rm N}\right)^2 = \left(r_{\rm N}\right)^2 - \left(d_{\sigma}\right)^2$ has been used. 

The expression (\ref{appendix_15}) has to be integrated over the time variable. To apply the advanced integration methods developed by \cite{Kopeikin1997}, we have to transform
(\ref{appendix_15}) from $\left(c t, \ve{x}\right)$ into terms of two new variables, $c \tau = \ve{\sigma} \cdot \ve{x}_{\rm N}$ and $\xi^i = P^{ij}\,x_{\rm N}^j$, which are independent of each other,
and obtain (note that after differentiation are performed, $\ve{\xi} = \ve{d}_{\sigma}$, hence $\left(d_{\sigma}\right)^2 = \ve{\xi} \cdot \ve{\xi} = \xi^2$)
\begin{equation}
	\frac{\ddot{\ve{x}}_{\rm g1} \left(t\right)}{c^2} = + \frac{4 G}{c^3} \sum\limits_{l=0}^{\infty} \frac{\left(-1\right)^l}{l!} \hat{W}_L\,\widehat{\partial}_L
	\left(\frac{2}{\left(r_{\rm N}\right)^3} - \frac{3\;\left(\xi\right)^2}{\left(r_{\rm N}\right)^5} \right) \, \ve{\sigma}\;,
\label{appendix_20}
\end{equation}

\noindent
where the double-dot in (\ref{appendix_20}) means twice the total derivative with respect to variable $\tau$, and $r_{\rm N}$ is the absolute value of the unperturbed light ray 
in terms of these new variables (\ref{Unperturbed_Lightray_Absolute_2}). 
Let us note that the left-hand side in (\ref{appendix_20}) depends on variable of coordinate time, because in (\ref{appendix_20}) the differentiations have to be performed, 
and afterwards one has to replace $c \tau$ by $\ve{\sigma} \cdot \ve{x}_{\rm N}\left(t\right)$ and $P^{ij} \xi_j$ by $d^i_{\sigma}$; see text below Eq.~(\ref{tau_0}).
The  differential operator (\ref{appendix_20}) has been given by Eq.~(\ref{Transformation_Derivative_3}).
To get the coordinate velocity of the light signal, one has to integrate (\ref{appendix_20}) over variable $c \tau$ and obtains for the spatial components
\begin{eqnarray}
	\frac{\Delta \dot{x}^i_{\rm g1} \left(t\right)}{c} &=& - \frac{4 G}{c^3} \sum\limits_{l=0}^{\infty} \frac{\left(-1\right)^l}{l!} 
	\hat{W}_L\,\widehat{\partial}_L \,\frac{c \tau}{\left(r_{\rm N}\right)^3}\;\sigma^i
	\nonumber\\ 
	&=& + \frac{4 G}{c^3} \frac{\partial}{\partial c \tau}\sum\limits_{l=0}^{\infty} \frac{\left(-1\right)^l}{l!}\,\widehat{\partial}_L
	\,\frac{\hat{W}_L}{r_{\rm N}}\;\sigma^i\;.
\label{appendix_30}
\end{eqnarray}

\noindent
A similar calculation can be performed for the second gauge term (\ref{Term2}), which yields
\begin{eqnarray}
	\frac{\Delta \dot{x}^{i}_{\rm g2} \left(t\right)}{c} &=& 
	- \frac{4 G}{c^2} \frac{\partial}{\partial c \tau} \sum\limits_{l=0}^{\infty} \frac{\left(-1\right)^l}{l!}\,\widehat{\partial}_{iL}\,\frac{\hat{X}_{L}}{r_{\rm N}}
	\nonumber\\ 
	&& \hspace{-1.5cm} - \frac{4 G}{c^2} \frac{\partial}{\partial c \tau} \sum\limits_{l=1}^{\infty} \frac{\left(-1\right)^l}{l!}\,\widehat{\partial}_{L-1}
	\,\frac{\hat{Y}_{iL-1}}{r_{\rm N}}
	\nonumber\\ 
	&& \hspace{-1.5cm} - \frac{4 G}{c^2} \frac{\partial}{\partial c \tau} \sum\limits_{l=1}^{\infty} \frac{\left(-1\right)^l}{l!}\,\frac{l}{l+1}\,\epsilon_{iab}\, 
	\widehat{\partial}_{a L-1}\,\frac{\hat{Z}_{bL-1}}{r_{\rm N}}\,.
\label{appendix_35}
\end{eqnarray}

\noindent
The second integration over variable $c \tau$, from lower integration limit $c \tau_0$ to upper integration limit $c \tau_1$, can be performed immediately and yields 
\begin{eqnarray}
	\Delta x^i_{\rm g1} \left(t_1, t_0\right) &=& \Delta x^i_{\rm g1} \left(t_1\right) - \Delta x^i_{\rm g1} \left(t_0\right),
        \label{appendix_40_A}
        \\
	\Delta x^i_{\rm g2} \left(t_1, t_0\right) &=& \Delta x^i_{\rm g2} \left(t_1\right) - \Delta x^i_{\rm g2} \left(t_0\right),
        \label{appendix_45_A}
        \end{eqnarray}

	\noindent
	with 
	\begin{eqnarray}
	\Delta x^i_{\rm g1} \left(t\right) &=& + \frac{4 G}{c^3}\sum\limits_{l=0}^{\infty} \frac{\left(-1\right)^l}{l!}\,\widehat{\partial}_L
	\,\frac{\hat{W}_L}{r_{\rm N}}\;\sigma^i\;, 
        \label{appendix_40}
         \\
	 \Delta x^i_{\rm g2} \left(t\right) &=& - \frac{4 G}{c^2} \sum\limits_{l=0}^{\infty} \frac{\left(-1\right)^l}{l!}\,\widehat{\partial}_{iL}\,\frac{\hat{X}_{L}}{r_{\rm N}}
	 \nonumber\\ 
	 && \hspace{-1.5cm} - \frac{4 G}{c^2} \sum\limits_{l=1}^{\infty} \frac{\left(-1\right)^l}{l!}\,\widehat{\partial}_{L-1}\,\frac{\hat{Y}_{iL-1}}{r_{\rm N}}
	 \nonumber\\ 
	 && \hspace{-1.5cm} - \frac{4 G}{c^2} \epsilon_{iab} \sum\limits_{l=1}^{\infty} \frac{\left(-1\right)^l}{l!}\,\frac{\left(-1\right)^l}{l!}\, 
	 \widehat{\partial}_{a L-1}\,\frac{\hat{Z}_{bL-1}}{r_{\rm N}}\;.
	 \label{appendix_45}
\end{eqnarray}

\noindent 
The time-dependence of the gauge terms (\ref{appendix_40_A}) and (\ref{appendix_45_A}) are just via the spatial positions of the unperturbed light ray at the time of emission 
and reception, $\ve{x}_{\rm N}\left(t_0\right)$ and $\ve{x}_{\rm N}\left(t_1\right)$. In line with (\ref{Appendix_C_1}), the terms (\ref{appendix_40_A}) and (\ref{appendix_45_A}) 
are added together, so we arrive at  
\begin{equation}
	\Delta \ve{x}_{\rm gauge} \left(t_1, t_0\right) =
	\Delta \ve{x}_{\rm g1} \left(t_1, t_0\right) + \Delta \ve{x}_{\rm g2} \left(t_1, t_0\right). 
\label{appendix_47}
\end{equation}

\noindent 
By inserting the differential operator (\ref{Transformation_Derivative_3}) into these solutions (\ref{appendix_40}) and (\ref{appendix_45}) one finds that  
these terms in (\ref{appendix_40}) and (\ref{appendix_45}) vanish separately in the asymptotic limit, 
\begin{eqnarray}
	\lim_{\tau = \pm \infty} \Delta \ve{x}_{\rm g1} \left(\ve{x}_{\rm N}\left(t\right)\right) &=& 0\,,
\label{appendix_50}
\\
	\lim_{\tau = \pm \infty} \Delta \ve{x}_{\rm g2} \left(\ve{x}_{\rm N}\left(t\right)\right) &=& 0\,, 
\label{appendix_51}
\end{eqnarray}

\noindent
with $c \tau = \ve{\sigma} \cdot \ve{x}_{\rm N}\left(t\right)$. Eqs.~(\ref{appendix_50}) and (\ref{appendix_51}) imply for (\ref{appendix_47}) the limits 
\begin{eqnarray}
        \lim_{\tau = \tau_0 \rightarrow - \infty \atop \tau = \tau_1 \rightarrow + \infty}  \Delta \ve{x}_{\rm gauge} \left(t_1, t_0\right) = 0\;, 
	\label{appendix_52}
\end{eqnarray}

\noindent 
which is more general that the asserted relation in (\ref{second_integration_gauge}).

\section{Proof of Eqs.~(\ref{Shapiro_minus_infinity}) and (\ref{Shapiro_plus_infinity})}\label{Appendix2}

In this appendix we will show relations (\ref{Shapiro_minus_infinity}) and (\ref{Shapiro_plus_infinity}). 
That means, we consider the limits  
\begin{eqnarray}
\lim_{\tau = \tau_0 \rightarrow - \infty \atop \tau = \tau_1 \rightarrow + \infty} \,\widehat{\partial}_{L} \,\ln \left(r_{\rm N} + c \tau\right),  
        \label{Appendix2_5}
\end{eqnarray}

\noindent 
where $r_{\rm N} = \sqrt{\xi^2 + c^2 \tau^2}$ with $\xi^2 = \ve{\xi} \cdot \ve{\xi}$, while the differential operator is given by Eq.~(\ref{Transformation_Derivative_3}), 
and $c \tau_0$ and $c \tau_1$ are defined by Eqs.~(\ref{sublabel_tau_0}) and (\ref{sublabel_tau_1}), respectively. 

First of all, we consider the first derivative of the logarithm in (\ref{Appendix2_5}) with respect to variable $c \tau$, 
\begin{eqnarray}
	\lim_{\tau = \tau_0 \rightarrow - \infty \atop \tau = \tau_1 \rightarrow + \infty} \, \frac{\partial}{\partial c \tau}\,\ln \left(r_{\rm N} + c \tau\right) &=& 
	\lim_{\tau = \tau_0 \rightarrow - \infty \atop \tau = \tau_1 \rightarrow + \infty} \,\frac{1}{\sqrt{\xi^2 + c^2 \tau^2}} = 0\;.
	\nonumber\\ 
        \label{Appendix2_10}
\end{eqnarray}

\noindent
Clearly, any further derivative of this term, with respect to either variable $\ve{\xi}$ or $c \tau$, increases the inverse power of $c \tau$ at least by one order. Therefore, 
all those terms in the differential operator (\ref{Transformation_Derivative_3}) which contain at least one derivative with respect to variable $c \tau$ will vanish in these limits. 
Hence, one has only to consider the term with $p=0$ of the differential operator in Eq.~(\ref{Transformation_Derivative_3}), given by
\begin{eqnarray}
        \widehat{\partial}^{\;\; p=0}_{L} &=& {\rm STF}_{i_1 \dots i_l}\;P_{i_{1}}^{j_{1}}\;...\;P_{i_l}^{j_l} 
        \,\frac{\partial}{\partial \xi^{j_{1}}}\;...\;\frac{\partial}{\partial \xi^{j_{l}}}\;, 
        \label{Appendix2_25}
\end{eqnarray}

\noindent
in line with the comment in the text below Eq.~(\ref{Shapiro_plus_infinity}). 

The logarithm in (\ref{Appendix2_5}) can be written in the form
\begin{eqnarray}
	\ln \left(r_{\rm N} + c \tau\right) &=& \ln \left|c \tau\right| + \ln \left(\sqrt{1 + \xi^2/c^2 \tau^2} \pm 1\right), 
        \label{Appendix2_15}
\end{eqnarray}

\noindent
where the plus sign in the argument of the logarithm is for $c \tau > 0$, while the minus sign in the argument of logarithm is for $c \tau < 0$. 
The first logarithm on the right-hand side of (\ref{Appendix2_15}) can be omitted, because of
\begin{eqnarray}
        && \hspace{-0.75cm} \frac{\partial}{\partial \xi^{j_{p+1}}}\;...\;\frac{\partial}{\partial \xi^{j_{l}}}\;\ln \left|c \tau\right| = 0\;. 
        \label{Appendix2_16}
\end{eqnarray}

\noindent
Thus, one only needs to consider the second logarithmic term on the right-hand side of (\ref{Appendix2_15}). 
In order to evaluate these limits in (\ref{Appendix2_5}), it is appropriate to introduce the dimensionless three-vector,  
\begin{eqnarray}
	\zeta^a &=& \frac{\xi^a}{c \tau} \quad \rightarrow \quad \frac{\partial}{\partial \xi^a} = \frac{1}{c \tau}\,\frac{\partial}{\partial \zeta^a}\;, 
       \label{Appendix2_30}
\end{eqnarray}

\noindent 
where the absolute value of this three-vector, $\zeta = \sqrt{\ve{\zeta} \cdot \ve{\zeta}}$, is a small quantity in the limits $c \tau \rightarrow \pm \infty$, that means 
$\zeta \rightarrow 0$, because the impact vector, $\ve{\xi}$, remains constant; see also Fig.~\ref{Diagram2}. 
The differential operator in (\ref{Appendix2_25}) in terms of three-vector $\zeta^a$ transforms into  
\begin{eqnarray}
        \widehat{\partial}^{\;p=0}_{L} &=& \frac{1}{\left(c \tau\right)^l}\,{\rm STF}_{i_1 \dots i_l}\;P_{i_{1}}^{j_{1}}\;...\;P_{i_l}^{j_l} 
        \,\frac{\partial}{\partial \zeta^{j_{1}}}\;...\;\frac{\partial}{\partial \zeta^{j_{l}}}\,. 
        \label{Appendix2_40}
\end{eqnarray}

\noindent 
Let us consider the term with the plus sign in (\ref{Appendix2_5}), which in terms of three-vector (\ref{Appendix2_30}) reads  
\begin{eqnarray}
	&& \lim_{\tau = \tau_1 \rightarrow + \infty} \,\widehat{\partial}_{L} \,\ln \left(r_{\rm N} + c \tau\right) 
	= \lim_{\tau = \tau_1 \rightarrow + \infty} \frac{1}{\left(c \tau\right)^l}
	\nonumber\\
	&& \times\, {\rm STF}_{i_1 \dots i_l} P_{i_{1}}^{j_{1}}\;...\;P_{i_l}^{j_l}\, 
	\frac{\partial}{\partial \zeta^{j_{1}}}\;...\;\frac{\partial}{\partial \zeta^{j_{l}}}\, \ln \left(\sqrt{1 + \zeta^2} + 1\right).  
	\nonumber\\
       \label{Appendix2_35}
\end{eqnarray}

\noindent 
The term in the second line of (\ref{Appendix2_35}) is finite, even for any values of $\zeta^2$. 
Thus, in view of the prefactor $(c \tau)^{-l}$, the term (\ref{Appendix2_35}) vanishes in the limit $c \tau \rightarrow + \infty$ and one obtains  
\begin{eqnarray}
        \lim_{\tau = \tau_1 \rightarrow + \infty} \, 
        \widehat{\partial}_{L} \,\ln \left(r_{\rm N} + c \tau\right) &=& 0\;. 
       \label{Appendix_Shapiro_plus_infinity}
\end{eqnarray}

\noindent 
Let us consider the term with the minus sign in (\ref{Appendix2_5}), which in terms of three-vector (\ref{Appendix2_30}) reads
\begin{eqnarray}
        && \lim_{\tau = \tau_0 \rightarrow - \infty} \,\widehat{\partial}_{L} \,\ln \left(r_{\rm N} + c \tau\right) 
        = \lim_{\tau = \tau_0 \rightarrow - \infty} \frac{1}{\left(c \tau\right)^l}
        \nonumber\\
        && \times\, {\rm STF}_{i_1 \dots i_l} P_{i_{1}}^{j_{1}}\;...\;P_{i_l}^{j_l}\, 
        \frac{\partial}{\partial \zeta^{j_{1}}}\;...\;\frac{\partial}{\partial \zeta^{j_{l}}}\, \ln \left(\sqrt{1 + \zeta^2} - 1\right). 
        \nonumber\\
       \label{Appendix2_50}
\end{eqnarray}

\noindent
The term in the second line of (\ref{Appendix2_50}) diverges in the limit $\zeta \rightarrow 0$. To determine this limit, a series expansion of the logarithm in (\ref{Appendix2_50}) 
is performed for $\zeta \ll 1$, which reads 
\begin{eqnarray}
	\ln \left(\sqrt{1 + \zeta^2} - 1\right) &=& - \ln \left(2 \right) + \ln \left(\zeta^2\right) - \frac{\zeta^2}{4} + {\cal O}\left(\zeta^4\right).
	\nonumber\\ 
        \label{Appendix2_60}
\end{eqnarray}

\noindent 
The constant $\ln (2)$ does not contribute, because a derivative of a constant with respect to variable $\zeta$ in (\ref{Appendix2_50}) vanishes. Similarly, the third term of power 
$\zeta^2$ and also terms of higher powers ${\cal O}\left(\zeta^4\right)$ on the right-hand side in (\ref{Appendix2_60}) vanish in the limit $\zeta \rightarrow 0$ and, in addition, also 
in view of the prefactor $(c \tau)^{-l}$ in (\ref{Appendix2_50}). Thus, only the logarithmic term remains, keeping in mind that the differential operator (\ref{Appendix2_40}) 
is acting on this logarithm. By transforming (\ref{Appendix2_60}) from three-vector $\ve{\zeta}$ back into three-vector $\ve{\xi}$ and by accounting for the statements below 
Eq.~(\ref{Appendix2_60}), one arrives at 
\begin{eqnarray}
	\lim_{\tau = \tau_0 \rightarrow - \infty} \, 
        \widehat{\partial}_{L} \,\ln \left(r_{\rm N} + c \tau\right) &=& 2\,\widehat{\partial}^{\;p=0}_{L} \ln \left| \ve{\xi}\right|.
        \label{Appendix_Shapiro_minus_infinity}
\end{eqnarray}

\noindent
By relations (\ref{Appendix_Shapiro_plus_infinity}) and (\ref{Appendix_Shapiro_minus_infinity}) the validity of Eqs.~(\ref{Shapiro_minus_infinity}) and (\ref{Shapiro_plus_infinity}) 
has been shown. Let us notice again, that the differential operator $\hat{\partial}_{L}$ on the left-hand side in (\ref{Appendix_Shapiro_plus_infinity}) and 
(\ref{Appendix_Shapiro_minus_infinity}) is given by Eq.~(\ref{Transformation_Derivative_3}), while on the right-hand side the 
differential operator $\widehat{\partial}^{\;p=0}_{L}$ is given by (\ref{Appendix2_25}). However, one may keep the differential operator $\hat{\partial}_{L}$ on the 
right-hand side in (\ref{Appendix_Shapiro_plus_infinity}) and (\ref{Appendix_Shapiro_minus_infinity}), because derivatives with respect to variable $c \tau$ in the 
differential operator (\ref{Transformation_Derivative_3}) would not contribute anyway.


\end{document}